\documentclass[12pt,article]{JHEP3}
\usepackage{amsmath,amssymb,euscript,array,mathrsfs,epsfig}
\usepackage[small]{caption}
\newcommand{\startappendix}{
\setcounter{section}{0}
\renewcommand{\thesection}{\Alph{section}}}
\newcommand{\Appendix}[1]{
\refstepcounter{section}
\begin{flushleft}
{\large\bf Appendix \thesection: #1}
\end{flushleft}}
\def\ben
{\begin{equation}}
\def\een{\end{equation}}

    \let\e=\varepsilon

\def\be{\begin{equation}}
\def\ee{\end{equation}}
\def\ba{\begin{array}}
\def\ea{\end{array}}

\def\dalemb#1#2{{\vbox{\hrule height .#2pt
        \hbox{\vrule width.#2pt height#1pt \kern#1pt
                \vrule width.#2pt}
        \hrule height.#2pt}}}

\newcommand{\bea}{\begin{eqnarray}}
\newcommand{\eea}{\end{eqnarray}}

\newcommand{\Tr}{{\rm Tr}\,}

\title{Weak coupling large-$\boldsymbol N$ transitions at finite baryon density}
\author{Timothy J. Hollowood and  S. Prem Kumar\\\\
{\it 
 Department of Physics, \\
 Swansea University, \\ 
Singleton Park,\\
 Swansea, SA2 8PP, U.K.  
}\\
E-mail: \email{t.hollowood, s.p.kumar@swansea.ac.uk}
} 
\author{Joyce C. Myers\\\\
{\it
Centre for Theoretical Physics,\\
University of Groningen,\\
Nijenborgh 4, \\
9747 AG, Groningen, The Netherlands.
}\\
E-mail: \email{j.c.myers@rug.nl}
}

\abstract{
We study thermodynamics of free $SU(N)$ gauge theory with a large number of colours and flavours on a three-sphere, in the presence of a baryon number chemical potential. Reducing the system to a holomorphic large-$N$ matrix integral, paying specific attention to theories with scalar flavours (squarks), we identify novel third-order deconfining phase transitions as a function of the chemical potential. These transitions in the {\em complex} large-$N$ saddle point configurations are interpreted as ``melting'' of baryons into (s)quarks. They are triggered by the exponentially large ($\sim e^N$) degeneracy of light baryon-like states, which include ordinary baryons, adjoint-baryons and baryons made from different spherical harmonics of flavour fields on the three-sphere. The phase diagram of theories with scalar flavours terminates at a phase boundary where baryon number diverges, representing the onset of Bose condensation of squarks.
}
\begin{document}
\section{Introduction and summary}
Unravelling the thermodynamic phase structure of gauge theories such as quantum chromodynamics (QCD) poses several challenging and interesting questions. The behaviour of cold, dense matter in QCD is particularly elusive due to the so-called sign problem, and because low temperature, finite density transitions are expected to occur in the regime of strong gauge coupling \cite{Stephanov:2007fk}. The large-$N$ expansion promises to be a useful means of understanding certain aspects of the relevant physics. In recent years, the 't Hooft and Veneziano large-$N$ limits \cite{largen1,largen2} have been applied to argue the existence of a low temperature quarkyonic phase \cite{McLerran:2007qj, Hidaka:2008yy} where baryons condense. A second line of investigation of large-$N$ limits of QCD-like theories, involves the use of tools provided by gauge/gravity dualities \cite{maldacena, magoo} to describe strongly interacting theories with a large number of flavours \cite{Nunez:2010sf,Bigazzi:2011it}.

Drawing motivation from these research directions, in this paper we study the phase diagram of $SU(N)$ gauge theories with both fundamental scalars and fermions in the Veneziano large-$N$ limit, on a spatial three-sphere,  and with the gauge coupling set to zero. 
For asymptotically free theories, this is a natural limit to consider when the radius of the sphere is much smaller than the dynamical length scale at which gauge interactions become strong. Despite being in a compact space, a non-trivial phase structure is possible because of the large-$N$ limit which plays the role of the thermodynamic limit.
 
It is now well appreciated, following the works of \cite{Sundborg:1999ue, Aharony:2003sx}, that even free large-$N$ theories on compact spaces can have a non-trivial phase structure which may, in certain cases, be continued through to the interacting situation. For theories with adjoint matter, most famously for the free ${\cal N}=4$ supersymmetric Yang-Mills (SYM) theory on $S^3$, this study reveals a Hagedorn density of states leading to a first order deconfinement phase transition with increasing 
temperature\footnote{While it is not known whether this first order transition in ${\cal N}=4$ SYM continues to exist for small non-zero `t Hooft coupling, the corresponding weak coupling calculation was carried out for pure $SU(N)$ Yang-Mills theory (at large $N$) and the first order deconfinement transition found to persist \cite{Aharony:2005bq}.}.  The deconfinement transition survives the continuation to strong 't Hooft coupling wherein the gravity dual displays a Hawking-Page transition from thermal AdS space to an AdS black hole geometry \cite{witten1}. The qualitative agreement remarkably persists upon inclusion of chemical potentials for R-symmetry charges in ${\cal N}=4$ SYM \cite{yamada, critical, gubser}. 

We consider models containing both fundamental scalars (``squarks'') and fermions (``quarks'') as this is the generic matter content of supersymmetric theories that arise in holographic descriptions of gauge theories with fundamental flavour fields (e.g. \cite{Casero:2006pt} and the D3-D7 system \cite{Nunez:2010sf, karchkatz}). Thermal transitions (at zero density) in such setups at zero and weak couplings were considered in \cite{Schnitzer:2004qt, Schnitzer:2006xz, Basu:2008uc, Skagerstam:1983gv}. The crucial new feature upon introducing a chemical potential for (s)quark number or baryon number, is that the effective action becomes complex due to the breaking of charge conjugation symmetry. The thermodynamics of gauge theory on $S^3$ is encoded in the effective action for the Polyakov loop matrix $U$, or the holonomy of the gauge field around the Euclidean thermal circle. This effective theory on a compact space yields a matrix model for $U$. The effective action being complex, it turns out that the matrix integral must be interpreted in a certain holomorphic sense so that the integral picks up, and is dominated by, complex saddle point configurations at large-$N$ 
(see e.g. \cite{Dijkgraaf:2002vw}). This was first shown in 
\cite{Hands:2010zp} where theories with fundamental fermions were studied (see also \cite{Hands:2010vw, Azakov:1986pn}). In this paper we will use the same techniques revealing a wider range of phenomena that occur when fundamental scalars are included. We further clarify the physical reasons behind these and also some of the results found in \cite{Hands:2010zp}.

\begin{figure}[h]
\begin{center}
\epsfig{file=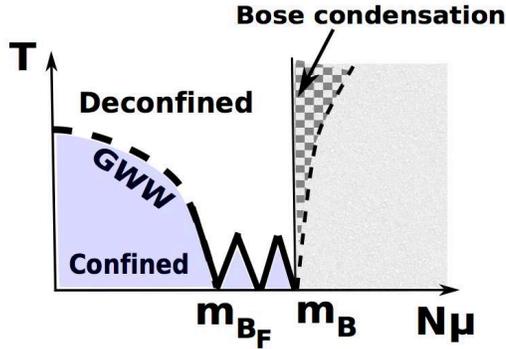, width =3in}
\end{center}
\caption{\small{ General features of the phase diagram.  Thick black lines  correspond to third order phase transitions.  Sawtooth features are associated to the filling of low lying Fermi levels.
The first such feature appears on either side of
$N\mu=m_{B_F}$, the mass of the lightest baryon made from fermions (quarks). 
The value $N\mu=m_B$, the mass of the lightest baryon constituted of scalar modes, represents the onset of Bose condensation of squarks and is the boundary of the phase diagram for the free theory. With interaction potentials typical of supersymmetric models,  the theory for $N\mu> m_B$  will enter an unstable Higgs phase (shaded region), along with a locally stable deconfined phase (thatched region).
}}
\label{phasediag1} 
 \end{figure}
Our results are summarized in Figure\eqref{phasediag1}, which provides a general picture of the phase diagram for these theories. The theory at $\mu=0$, with increasing temperature,  experiences a third order Gross-Witten-Wadia (GWW) \cite{Gross:1980he, Wadia:1980cp} phase transition characterized by the formation of a gap in the large-$N$ eigenvalue distribution of the unitary matrix $U$. The non-analyticity at $\mu=0$  was first noted in \cite{Schnitzer:2004qt}.
Although there is no natural order parameter for confinement at finite $N$ in the presence of fundamental matter fields, non-analytic behaviour in the Polyakov loops occurs in the large-$N$ Veneziano limit and can be interpreted as deconfinement of mesons to quarks. 

The new features in the phase diagram, at low temperature and non-zero chemical potential, can be explained as follows. By Gauss's law on $S^3$, the grand canonical partition function must be thought of as a sum over gauge-singlet states only. A positive chemical potential $\mu >0$ essentially favours baryons over anti-baryons. 
There are primarily three distinct types of baryonic states whose masses and degeneracies determine the nature of the phase diagram: 
\begin{itemize}
\item{ Ordinary baryons $B_F$ constituted of fundamental fermions or quarks.}
\item{ Baryon operators $B_S$, composed of fundamental scalars and in particular,  various spherical harmonics of squark fields on $S^3$.}
\item{The third type of states which will be relevant for our discussion are the ``adjoint-baryons'', $B_{\rm Ad}$.  Adjoint baryon operators involve any number of insertions of adjoint fields into an ordinary baryon operator. The adjoint fields could be harmonics of the gauge field on $S^3$ or the modes of any other adjoint matter field in the theory. }
\end{itemize}

At large $N$, the degeneracy of all such states grows exponentially with $N$, as we will see below (for related discussions see \cite{Hidaka:2008yy,Hanany:2008sb}). This exponential growth of baryon states has a very important effect: although baryons are exponentially suppressed due to their masses being  ${\cal O}(N)$, their exponentially large degeneracies result in a competition between energy and entropy, wherein the latter can overwhelm the former in the grand canonical ensemble. Consequently, for any fixed low temperature we may argue the existence of a phase boundary at a value of $\mu$ determined by the mass and the degeneracy of the lightest baryon state on $S^3$. 

The expectations above are confirmed and complemented beautifully by the explicit behaviour of the {\em complex} saddle points of the large-$N$ matrix model for $U$. As $\mu$ approaches the mass of the lightest baryon state, and precisely when the entropy factor overwhelms the Boltzmann suppression, the distribution of eigenvalues of $U$ along a closed contour in the complex plane, experiences a ``gapping'' transition of the GWW type. This is accompanied by a smooth increase in baryon number, and the phase transition is found to be of third order.  The scaling of the grand potential with the number of flavours indicates that the confined phase should be attributed to an ensemble of gauge-singlet mesons and in this sense the quarks are confined.
Across the gapping transitions we find that gauge-invariant order parameters holomorphic in $U$, namely $\langle\tfrac{1}{N}\Tr\,U^n\rangle$, acquire non-zero values. Non-vanishing expectation values for Polyakov loop observables are expected in a deconfined phase \cite{svetitsky}. For this reason we term the non-analyticities as deconfinement transitions across which baryons ``melt''. The anti-holomorphic observables,  $\langle\tfrac{1}{N}\Tr\,U^{\dagger n}\rangle$ remain non-vanishing in all phases reflecting the asymmetry under charge conjugation in the presence of a (positive) chemical potential.

If the lightest state(s) happen to be ordinary baryons $B_{F}$, composed of light quark modes, two successive third order transitions occur on either side of the baryon mass at  $N\mu= m_{B_F}\mp T\,\ln\,\mathfrak{D}(B_F)$. Here ${\mathfrak D}$ is the degeneracy of the light baryon state.
 The first is a gapping transition followed immediately by a transition back to the ungapped or confined phase, with increasing $\mu$. Across the two transitions, the lowest lying Fermi levels get filled and the baryon number increases by a discrete step. In between two such transitions, a `spike' of deconfined phase penetrates all the way down to zero temperature. These are the features shown in Figure\eqref{phasediag1}.

In theories with squarks, an additional feature appears in the phase diagram. For baryons made from scalar flavours, complete antisymmetry of colour indices means that the lightest baryon cannot always be constituted solely of the lightest squark harmonics on $S^3$ (the zero modes). When the number of squark flavours is smaller than the number of colours, non-trivial baryon operators $B_S$ can exist only if higher harmonics of the squark fields are also included. Alternatively, one may form non-vanishing baryon operators $B_{\rm Ad}$ by insertions of adjoint fields. In all cases, our matrix model calculation explicitly yields a third order deconfining transition at $N\mu= m_{B} - T\ln\,{\mathfrak D(B)}$, where $B$ is the lightest baryon state composed of scalar flavours and ${\mathfrak D}(B)$, the associated degeneracy. Beyond this transition, we find that the baryon number grows without bound and eventually diverges at $N\mu = m_B$ . We identify this as the Bose condensation of squarks/baryons and the phase diagram cannot be continued past this boundary in the free theory. 

The onset of Bose condensation of squarks is accompanied by the collapse of the  distribution of eigenvalues of 
$U$ to a point, so that $U=1$. This marks the entry of the field theory into a ``Higgs phase'' where gauge-invariant operators carrying (positive) baryon number could acquire vacuum expectation values (VEVs). It is not possible to see this in the non-interacting theory.
 If interactions, such as those in supersymmetric QCD-like 
 theories\footnote{Note that the formulation of rigidly supersymmetric, non-conformal gauge theories on curved spaces, such as $S^3\times{\mathbb R}$ is subtle as it requires the introduction of new terms in the action \cite{Sen:1985ph,Festuccia:2011ws,Jia:2011hw}
}, were to be switched on, flat directions in the scalar potential would lead to runaway, unstable VEVs for large enough baryon number chemical potential \cite{yamada, critical, Chen:2009kx}. At any finite temperature however, there would be a metastable phase (thatched region in Figure\eqref{phasediag1}) of deconfined (s)quarks at the origin of the Higgs branch \cite{yamada, critical}.  In the $\mu-T$ plane, the width of the metastable region, at weak coupling, is determined by the thermal mass of the squark modes.

Our motivation for exploring the large-$N$ phase structure above is to glean information on the possible thermodynamic behaviour of such theories, which could be used as a point of reference for investigating similar theories at strong coupling within the framework of gauge/string dualities \cite{inprep}. While it may be unreasonable to expect a putative strongly coupled gravity dual to reproduce the features in Figure\eqref{phasediag1}, certain aspects are possibly robust enough to exist at any coupling. The exponential (in $N$) growth of baryonic states appears unlikely to be changed by interactions; baryon-baryon and meson-baryon interactions are intrinsically large, involving powers of $N$ but not exponentials \cite{Dashen:1993as,Luty:1994ua}. The cold deconfined phase at finite density  in the phase diagram suggests a dual, extremal black hole geometry at strong coupling.

The organization of the paper is as follows: In Section 2, we introduce notation and summarize the matter content of the theories in question. Section 3 is devoted to the basic features/transitions in the holomorphic large-$N$ matrix model, obtained from the theory truncated to lightest modes. We also show how the observed non-analyticities coincide with the expectations from counting of baryon operators. In Section 4, we show that in theories with a small number of squark flavours, it becomes necessary to include higher harmonics and heavier modes to obtain the correct phase diagram. In Section 5 we review the $\mu=0$, finite $T$ behaviour of the partition function. We end with Section 6, a summary of the phase diagram and possible lessons for related systems at strong coupling.

\section{Free gauge theory on $S^3$}
To evaluate the grand canonical partition function of  free gauge theories, we use the approach of $\cite{Aharony:2003sx}$. 
We consider $SU(N)$ gauge theory with $\nu_F N$ Dirac fermions (``quarks''), and 
$\nu_S N$ complex scalars (``squarks''), all transforming in the fundamental representation of the gauge group. The number of flavours is chosen to naturally scale with $N$. The theory with $\nu_S=0$ has already been studied in \cite{Hands:2010zp}. Below we extend this to include fundamental scalars, and further provide a natural interpretation of the ensuing phase diagram as a function of temperature and baryonic or quark chemical potential. The thermal partition function of the gauge theory on $S^3$ is computed by formulating the field theory in Euclidean signature, on $S^3\times S^1$ with the $S^1$  circumference $\beta = \frac{1}{T}$, playing the role of the inverse temperature.

For the free field theory on $S^3$, the partition sum follows from one loop fluctuation determinants for the Kaluza-Klein harmonics of all fields on $S^3$ subject to the constraint from Gauss' law. This latter constraint is necessary since the sum must be over physical, gauge-invariant states, and in practice, it is elegantly implemented by introducing a background thermal Wilson line for the zero mode of the gauge field on $S^3$: 
\bea
&&U = e^{i\alpha}\,\,\nonumber\\
&&
\alpha\,\equiv\, \frac{1}{{\rm Vol}(S^3)}\,
\oint_{S^1\times S^3}\, A_0  \,= \,{\rm diag}(\theta_1,\theta_2,\ldots, \theta_N)\,,\qquad 
\sum \theta_a =0 \,\,{\rm mod}\,\, 2\pi\,. 
\label{eval}
\eea
$SU(N)$ gauge transformations can be used to put this Wilson line - also known as the Polyakov loop, in a diagonal form as above. The problem of calculating the partition function is then reduced to determining the quantum effective action for the ``light'' zero modes above, after integrating out all Kaluza-Klein harmonics on $S^3$ as quadratic fluctuations.

We can consider the general situation where our 3+1 dimensional gauge theory has both adjoint and fundamental matter fields
\footnote{ This is a subtle issue once interactions are switched on and the gauge coupling acquires scale dependence. For an asymptotically free theory we must ensure that the radius of the $S^3$ is sufficiently small compared to the relevant strong coupling scale. If the theory is infrared (IR) free, then the $S^3$ must be taken to be large, and the flavour masses small compared to the scale of any putative Landau pole.}. Let $n_S$ and $n_F$ be the number of adjoint (real) scalars and chiral fermions respectively,
\bea
&&\phi_I\,,\qquad I = 1,2,\ldots n_S\,,\\\nonumber
&&\psi_A\,,\qquad A=1,2,\ldots n_F\,.
\eea
For simplicity we take the adjoint fermions and scalars to have vanishing  bare mass, with the scalars conformally coupled to the curvature of the three-sphere.
 
In addition, we take $2 N_S$ complex scalars (``squarks'') in the fundamental representation, each with mass $m_S$, and $N_F$ fundamental Dirac fermions (``quarks'') of mass $m_F$,
\bea
&& (q_i\,,\tilde q_i)\,: \qquad (N,\bar N)\,\,{\rm of}\,\,{SU(N)}\,\qquad
i=1,2,\dots N_S\\\nonumber
&& (\psi_i\,,\tilde\psi_i)\,: \qquad (N, \bar N)\,\, {\rm of}\,\,{SU(N)}\,\qquad i=1,2,\ldots N_F\,.
\eea
Our labelling of the flavour fields is motivated by supersymmetric theories with fundamental matter (SUSY QCD), although the analysis we present is general without any reference to supersymmetric theories. We also take all scalar flavours (``squarks'') to be {\em non-minimally} coupled to the curvature of $S^3$, which ensures that the fluctuations on $S^3$ have a gap even when the flavour masses are set to zero. In a conformal field theory (as is the case for certain supersymmetric examples), the scalars would be massless but conformally coupled to the curvature, ensuring a gap.

We define a $U(1)_B$ global symmetry, under which the fundamental fermions and scalars $(q_i,\psi_i)$ carry a charge $+1$, whilst the anti-fundamental fields $(\tilde q_i,\tilde\psi_i)$ have charge $-1$. We will refer to this as ``quark number'' or ``squark number''. The number of baryons is related to it by an extra factor of $1/N$.
Introducing a chemical potential for the squark or quark number, the effective action for the theory on a spatial $S^3$ takes the form of a unitary matrix model (see e.g. \cite{Aharony:2003sx, yamada, critical, Schnitzer:2004qt, Hands:2010zp}),
\be
{\cal Z}_{S^3\times S^1}\,=\, \int[dU]\, \exp\left(- S_{\rm eff}[U]\right)\,.
\ee
The effective action $S_{\rm eff}$ consists of two separate contributions, one from the adjoint sector and the other from the flavour sector, taking the general form
\bea
S_{\rm eff} \,=\, &&S_{\rm adj} + S_{\rm fund}\,\\\nonumber
=\,&&-\sum_{n=1}^\infty \left[{\cal A}_n\,\Tr\, U^n\,{\rm Tr}\,\, U^{\dagger n}\,+\, {\cal F}_n \,({\rm Tr}\, U^n\, e^{n \beta\mu} + {\rm Tr}\, U^{\dagger n}\, e^{-n \beta\mu}) \right]\,.
\eea
The adjoint fields are uncharged under $U(1)_B$ and the ensuing contributions are proportional to  ${\rm Tr}\,U^n\,{\rm Tr}\, U^{\dagger n}$. These terms are invariant under large gauge transformations that ``wind'' around the thermal $S^1$ up to an element of ${\mathbb Z}_N$. Fundamental flavour fields break the ${\mathbb Z}_N$ invariance explicitly and give rise to single-trace terms of the type ${\rm Tr}\,U^n$ and ${\rm Tr}\,U^{\dagger n}$\,. The most important point to note is that, due to the non-zero baryon number chemical potential, the effective action is {\em complex}. The coefficients ${\cal A}_n$ and ${\cal F}_n$ can be explicitly written in terms of single particle partition functions $z_F, z_S$ and $z_V$, for the fermion, scalar and vector harmonics respectively, on the three-sphere,
\bea
&& {\cal A}_n\,=\, \frac{1}{n}\big[z_V(x^n)+n_S\,z_S(x^n)- (-1)^n \,n_F\,z_F(x^n)\big]\\\nonumber\\\nonumber
&& x\,\equiv\, e^{-{1}/{T R}}\,\qquad z_{V}(x)\,\equiv\, \frac{6x^2-2x^3}{(1-x)^3}\qquad z_S(x)\,\equiv\, \frac{x(1+x)}{(1-x)^3}\qquad z_F(x)\,\equiv\, \frac{4 x^{3/2}}{(1-x)^3}\,.
\eea
Here $R$ is the radius of $S^3$.
The degeneracies $d_\ell$ and energies $\e_\ell$ 
of the different modes, which lead to each of the partition sums above are summarized in Table \eqref{table} in the Appendix. For the flavour modes the coefficients ${\cal F}_n$ are given as
\bea
{\cal F}_n\,=\, \frac{1}{n} \big[
N_S\,\,Z_S\left(n\tfrac{\beta}{R}, \,m_SR\right)\,  -  \,N_F\,\,(-1)^n\,Z_F\left(n\tfrac{\beta}{R}, \,m_F R\right) \big].
\eea
As before $Z_S$ and $Z_F$ are partition functions of scalar and fermion harmonics on $S^3$ with non-vanishing bare masses $m_S$ and $m_F$ respectively:
\bea
&& Z_S(\tfrac{\beta}{R},\,m_S R)\,=\,\sum_{\ell =0}^\infty 2 (\ell+1)^2 
\exp\left(-\frac{\beta}{R}\sqrt{\ell(\ell+2) + \Xi+m_S^2 R^2}\right)\\\nonumber
&&Z_F(\tfrac{\beta}{R},\,m_F R)\,=\,\sum_{\ell =1}^\infty 2 \ell (\ell+1) 
\exp\left(-\frac{\beta}{R}\sqrt{\left(\ell+\tfrac{1}{2}\right)^2+m_F^2 R^2}\right),
\eea
where $\Xi$ denotes the coupling to the Ricci scalar of $S^3$. For conformal coupling $\Xi=1$, while minimal coupling corresponds to $\Xi=0$.

The effective action in the adjoint sector scales as $N^2$, for large $N$ (taking $n_s\sim n_F \sim {\cal O}(1)$) as each term in the action is a double trace operator, whilst the effective action for the flavour modes scales as $N^2$ only if $N_F\sim N_S\sim{\cal O}(N)$,
\be
\nu_S\,\equiv\,\frac{2 N_S}{N}\,,\qquad \nu_F \,\equiv\,\frac{2 N_F}{N}\,.
\ee
It is worth repeating that in our notation, the theory has $2 N_S$ complex fundamental scalars, and $N_F$ Dirac fermions in the fundamental representation.
When $N_F=N_S$, the matter content in the flavour sector is that of $N_F$ hypermultiplets in the language of ${\cal N}=2$ supersymmetry. Similarly when $n_S=6$ and $n_F=4$, the adjoint sector has the matter content of the ${\cal N}=4$ theory.

\section{Low temperature large-$N$ transition}

At low temperatures $TR \ll 1$, the adjoint modes are exponentially suppressed and we may ignore $S_{\rm adj}$. 
The same is true for most of the flavour excitations except for those with energies comparable to the chemical potential $\mu$. We take $\mu$ to be positive without loss of generality and study the behaviour of the system as $\mu$ is increased from zero. 
\subsection{Truncated model: Light scalars $m_S\ll m_F$}
Taking the scalar modes to be strictly lighter than the fermions, for non-zero masses, we begin our low temperature analysis by keeping only the lightest mode carrying baryon number. The lightest scalar mode charged under $U(1)_B$ is the zero mode with $\ell=0$ and energy $\e_0 = \sqrt{\Xi R^{-2}+m_S^2}$, and we focus attention on the regime
\be
0 < \mu \simeq \e_0 = \sqrt{\Xi R^{-2}+m_S^2}\,,\qquad\qquad TR\ll 1\,. 
\ee
To be precise we work in the limit 
\be
|R(\mu-\e_0)|\to 0\,,\qquad TR \to 0\,\quad {\rm with}\quad 
\frac{|\mu-\e_0|}{T} = {\rm fixed}\,.
\ee
In this limit, the anti-squarks are Boltzmann suppressed by factors of $e^{-\beta(\e_0+\mu)}$ and can be ignored, as can all heavier fluctuations.
It is then consistent to truncate the effective action to the lightest squark mode alone ($U(1)_B$ charge +1) so that 
\be
S_{\rm eff}[U] \,\simeq\,   N\,\nu_S\,{\rm Tr}\,\ln(1-\zeta\,U)\,,\qquad\zeta\,\equiv\, e^{-\beta(\e_0-\mu)}\,,
\ee
with $\zeta$, the effective fugacity. 
Within this truncation, the effective action describes all gauge-invariant states made from the lightest flavour modes ($\ell=0$ scalar harmonics) on $S^3$.
In this regime it is clear that the effective matrix model for $U$ has a complex action, requiring a non-standard approach towards its large-$N$ solution. 
\subsection{ Holomorphic matrix integral }
Rewriting the matrix integral for $U$ in the eigenvalue basis
\eqref{eval}, and including the Jacobian for this transformation - the well known Vandermonde determinant, we obtain,
\be
S_{\rm eff}[\theta_a] = -\sum_{ab = 1}^N \ln\left|\sin\left(\frac{\theta_a-\theta_b}{2}\right)\right| +  N\,\nu_S\,\sum_{a=1}^N\ln(1-\zeta\, e^{i\theta_a}) + i\,{\cal Q}\, N\sum_{a=1}^N\theta_a\,.
\ee
Here we have explicitly implemented the condition $\sum_a\theta_a=0$,
via a Lagrange multiplier ${\cal Q}$, enforcing the requirement that $U$ has unit determinant.
In the large-$N$ limit, we expect that the integral over $\theta_a$ is dominated by a saddle point. However, since the integrand is complex, we must allow for {\em complex} saddle points. In particular, assuming that there exists such a saddle point which dominates the integral at large-$N$, the corresponding equation of motion is,
\be
\frac{1}{N}\sum_{a=1\,(\neq b)}^N\,\cot\left(\frac{\theta_b-\theta_a}{2}\right)
+ \nu_S\,\zeta\,i\,\frac{e^{i\theta_b}}{1-\zeta e^{i\theta_b}}=\,i{\cal Q}\,.
\label{saddle}
\ee
The equation has solutions only if  $\theta_a$ are complex.
To this end, we analytically continue the equation by introducing  holomorphic variables,
\be
{z_a} \,\equiv\, e^{i\theta_a}\,,\qquad{\rm with}\qquad
\sum_{a=1}^N\,\ln z_a\,=\,0\,.
\ee
In terms of these variables the saddle point configuration satisfies
\be
\frac{1}{N}\sum_{a=1\,(\neq b)}^N\,\frac{z_b+z_a}{z_b-z_a}+ \nu_S\,\zeta\,\frac{z_b}{1-\zeta {z_b}}\,=\,{\cal Q}\,.
\label{sp}
\ee
Summing over $b$, from this equation we learn that the Lagrange multiplier ${\cal Q}$ (times $N$) has the interpretation of average baryon number in the grand canonical ensemble,
\be
{\cal Q}\,=\, \nu_S\,\frac{1}{N}\sum_{a=1}^N\,\frac{\zeta\,z_a}{1-\zeta\,z_a}\,=\,\frac{T}{N^2} \frac{\partial\ln{\cal Z}}{\partial \mu}\,.
\label{number}
\ee
The squark number is $N^2 {\cal Q}$.
For low temperatures, when $\mu < \e_0$, the fugacity is exponentially small and the squark number should be suppressed. As $\mu$ is increased, we expect something interesting to happen since there is the possibility of a pole in the above expression for ${\cal Q}$ leading to a divergent baryon number, signalling the onset of Bose condensation.

\subsection{Low $\mu$ confined phase}
To solve the saddle point equation at large $N$, we take the eigenvalues $z_a$ to lie on a contour ${\cal C}$ in the complex $z$-plane. When the fugacity vanishes we know that ${\cal C}$ is the unit circle - this follows from the repulsive pairwise Vandermonde potential between the angular coordinates $\theta_a$. In the absence of any other potential terms at zero temperature and zero chemical potential, the eigenvalues spread uniformly around the unit circle. This is the confined phase of the theory. When the temperature is small, with a non-zero chemical potential and $\zeta < 1$, we look for a closed contour ${\cal C}$ which is an appropriate deformation of the unit circle. 

It is natural to introduce a parameter $t$, with $-\pi\leq t\leq \pi$ and treat the positions of the continuum of eigenvalues as a function $z(t)$. 
When the fugacity $\zeta$ vanishes and ${\cal C}$ is the unit circle we must have $z(t)=e^{it}$. More generally, we require $z(-\pi) = z(\pi)$, for ${\cal C}$ to be a closed contour. In this continuum limit all discrete sums can be replaced by integrals over ${\cal C}$ weighted by an eigenvalue density function $\rho(z)$:
\be
\frac{1}{N}\sum_{i=1}^N\cdots \, \to \int_{-\pi}^\pi \frac{dt}{2\pi}\cdots\,=\,\oint_{\cal C}\frac{dz}{2\pi i}\,\rho(z)\cdots\,\qquad \rho(z)\equiv i \frac{dt}{dz}\,.
\label{defrho}
\ee
The eigenvalue density satisfies two constraints -- a normalization condition (fixing the number of eigenvalues to $N$), and the unit determinant condition on $U$:
\be
\oint_{\cal C} \frac{dz}{2\pi i}\,\rho(z)\,=\,1\,,\qquad
\oint_{\cal C} \frac{dz}{2\pi i}\,\rho(z)\,\ln z\,=\,0\,.
\label{norm}
\ee
The second requirement actually fixes the orientation of the branch cut of $\ln z$, so that it extends along the negative real axis from the origin to 
$z=-\infty$. 

When the eigenvalues are uniformly distributed on the unit circle (the case with vanishing $\zeta$), $\rho(z)=\frac{1}{z}$. In this parametrization, the simple pole at the origin (enclosed by ${\cal C}$) ensures that the eigenvalue density is normalized correctly. As the fugacity is increased smoothly from zero, as long as ${\cal C}$ remains closed, we do not expect any further singularities to appear inside ${\cal C}$.
In the continuum limit the saddle point equation \eqref{sp} is
\be
{\cal P}\oint_{\cal C}\frac{dz'}{2\pi i}\,\,\rho(z')\,\,\frac{z+z'}{z-z'}\,\,+\,\,
\nu_S\,\zeta\,\frac{z}{1-\zeta\,z}\,=\,{\cal Q}\,,\qquad z\in{\cal C}\,,
\ee
where the integral over the contour ${\cal C}$ is to be treated as a principal value since the integrand has a singularity when $z'=z$ along the contour of integration.

It is fairly straightforward to see that, for small $\zeta$, $\rho(z)$ can be uniquely fixed to be,
\be
\boxed{
\rho(z)\, = \,\frac{1}{z}\, + \, \frac{\nu_S\,\zeta}{1-\zeta\,z}\,\,,\qquad {\cal Q} =0\,}\qquad \zeta\,\,{\rm small}\,.
\ee
Even without an explicit knowledge of ${\cal C}$ (which can now be deduced), the assumption of analyticity has  allowed us to determine $\rho$. 

The solution is valid as long as the simple pole at $z=\frac{1}{\zeta}$ lies outside ${\cal C}$. This is certainly true when $\zeta$ vanishes and should continue to be the correct picture as $\zeta$ is increased smoothly. However, we expect something interesting to happen as $\zeta$ is increased and the pole at 
$z=\frac{1}{\zeta}$, approaches the contour ${\cal C}$ on which the eigenvalues actually lie. Note also that, as long as this pole lies outside ${\cal C}$, the baryon number ${\cal Q}$ (defined in \eqref{number}) vanishes identically by Cauchy's theorem. 

Since $\rho(z)$ is fixed, we can now determine ${\cal C}$ using Eq.\eqref{defrho},
\be
e^{it} \,= \,\frac{z}{(1-\zeta z)^{\nu_S}}\,,\qquad -\pi < t \leq \pi\,
\label{implicitc}
\ee
which implicitly fixes $z$ as a function of $t$. As expected, when $\zeta$ vanishes, we recover the unit circle $z=e^{it}$.

In this phase the expectation values of all the ``holomorphic observables'', $\Tr \,U^n$ vanish identically by the residue theorem, whereas the operators $\Tr\, U^{\dagger n}$ are all non-vanishing,
\be
\langle\tfrac{1}{N}\Tr\,U^n\rangle\,=\,\oint_{\cal C}\frac{dz}{2\pi i}\,
\rho(z)\,z^n\,=0\,,\qquad
\langle\tfrac{1}{N}\Tr\,U^{\dagger n}\rangle\,=\,\oint_{\cal C}\frac{dz}{2\pi i}\,\frac{\rho(z)}{z^n}\,= \nu_S\,\zeta^n\ .
\label{polloops1}
\ee
This is a consequence of the difference between states carrying baryon and anti-baryon number (made from $\ell =0$ modes of the scalars). 
The baryon number chemical potential biases the system towards baryons as opposed to anti-baryons. 

\subsection{Phase transitions}
The form of the contour ${\cal C}$, and the location of the poles and zeroes of $\rho(z)$ relative to it, signal the possible onset of phase transitions as we show below. For general values of $\nu_S$ and $\zeta$, 
$\rho(z)$ has one simple pole and one simple zero in the $z$-plane,
\be
{\rm Pole:}\quad Z_\infty\,=\,\frac{1}{\zeta}\,,\qquad\qquad
{\rm Zero:}\quad Z_0\,=\,\frac{1}{\zeta\,(1- \nu_S)}\,.
\ee
Interestingly, both lie on the real axis. Now, there are two qualitatively distinct possibilities depending on whether $Z_0$ is positive or negative.
\begin{figure}[h]
\begin{center}
\epsfig{file=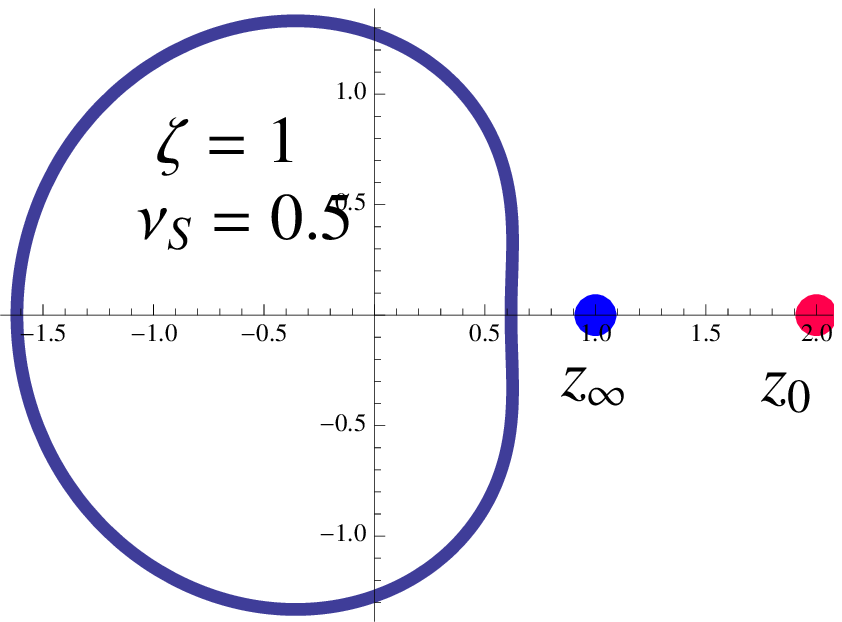, width =1.3in}
\hspace{0.1in}
\epsfig{file=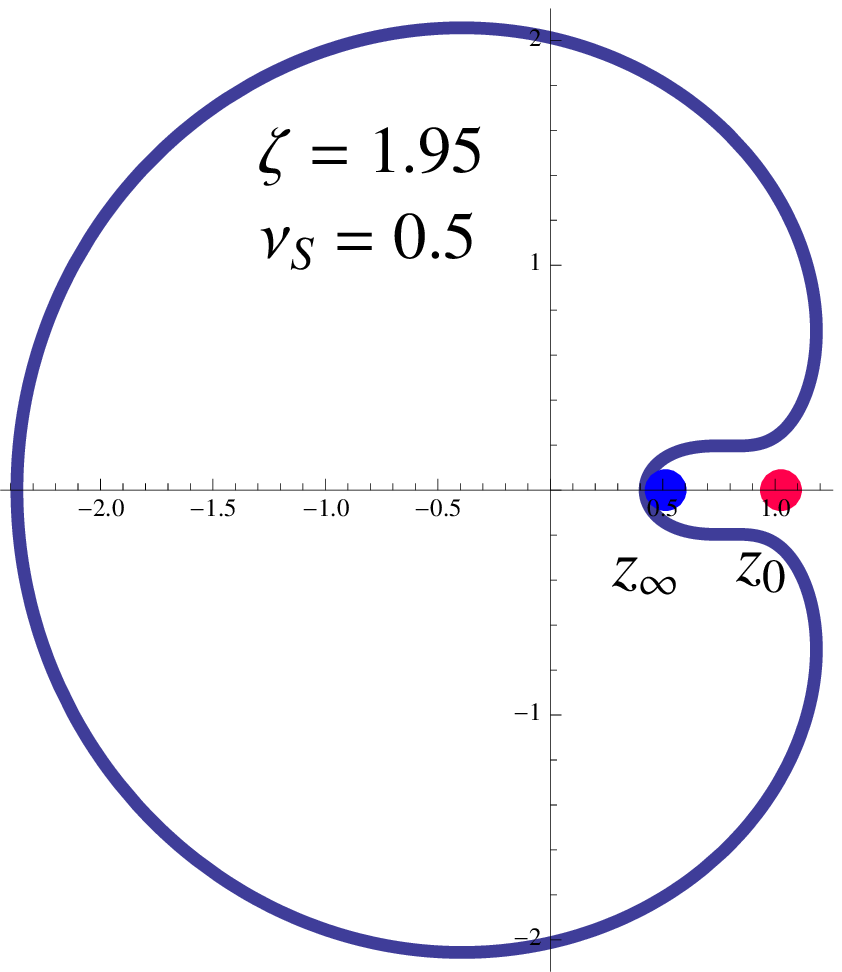, width =1.2in}
\hspace{0.2in}
\epsfig{file=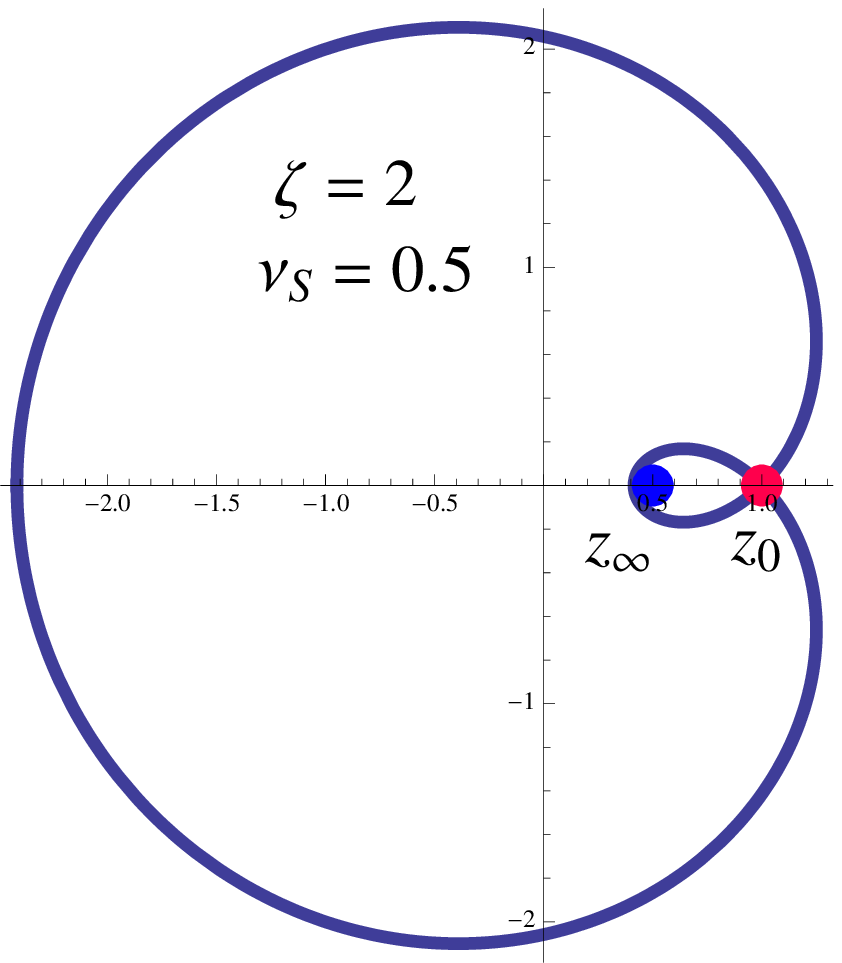, width =1.2in}
\hspace{0.2in}
\epsfig{file=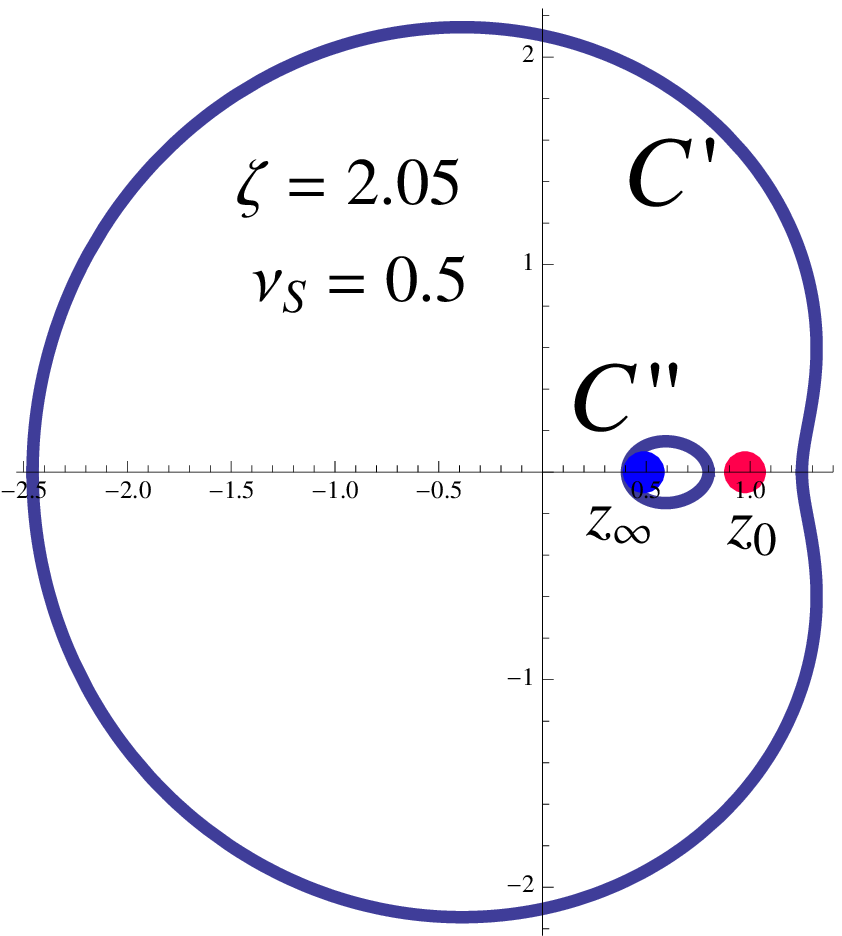, width =1.3in}
\end{center}
\caption{The relative positions of the contour ${\cal C}$ and the pole $Z_\infty$ and zero $Z_0$ with increasing $\zeta$  for $\nu_S < 1$. The pole repels the eigenvalue distribution, and eventually the zero hits the contour ${\cal C}$. Further increase in $\zeta$ causes the distribution to split into disconnected components ${\cal C}'$ and ${\cal C}''$.} 
 \label{contour} 
 \end{figure}
\subsubsection{ $2\boldsymbol N_S<\boldsymbol N\,$: splitting eigenvalue ``bubbles''  } 
Figure\eqref{contour} shows the contour ${\cal C}$, and the positions of the pole and zero as the fugacity (or the chemical potential) is increased. When ${\nu_S}<1$, $Z_0$ and $Z_{\infty}$ lie to the right of ${\cal C}$. However, as $\zeta$ is increased both the pole and the zero approach ${\cal C}$; the distribution of eigenvalues appears to be ``repelled'' by the pole, splitting the distribution in two disconnected contours when eventually the zero actually crosses the contour ${\cal C}$. This splitting process leads to a non-analyticity in the eigenvalue density, 
suggestive of a phase transition.

The critical value of $\zeta$ for the onset of this behaviour can be obtained by substituting $Z_0$ into the implicit expression \eqref{implicitc} for the contour. The requirement that $\zeta$ be real the leads to the solution,
\be
\zeta_{\rm crit}= \frac{(1-\nu_S)^{\nu_S-1}}{(\nu_S)^{\nu_S}}\,,\qquad\qquad t = \pm\, \nu_S \pi\,,\qquad \nu_S < 1\,.
\ee
Therefore there are two values of the parameter $t$ ($=\pm \,\nu_S\pi$), which simultaneously map to the point $Z_0$ when $\zeta=\zeta_{\rm crit}$.
Hence, two points on the contour ${\cal C}$ come together at $z=Z_0$, when  the fugacity is tuned to the critical value.
 This explains the behaviour of the contour ${\cal C}$ encountered in Figure \eqref{contour}. In particular,
\bea
&&{\cal C}''\,=\,\{ z(t)\,;\, - \nu_S\pi \leq t < \nu_S\pi\}\\\nonumber
&&{\cal C}'\,=\,\{z(t)\,;\, -\pi < t \leq -\nu_S\pi\,, \,\nu_S\pi < t\leq \pi\}\,.
\eea
We infer from the extents of the two distributions as a function of $t$, that a fraction $\nu_S$ of the eigenvalues ($2 N_S$ out of  a total number $N$) have peeled off from the original contour ${\cal C}$. 

\begin{figure}[h]
\begin{center}
\epsfig{file=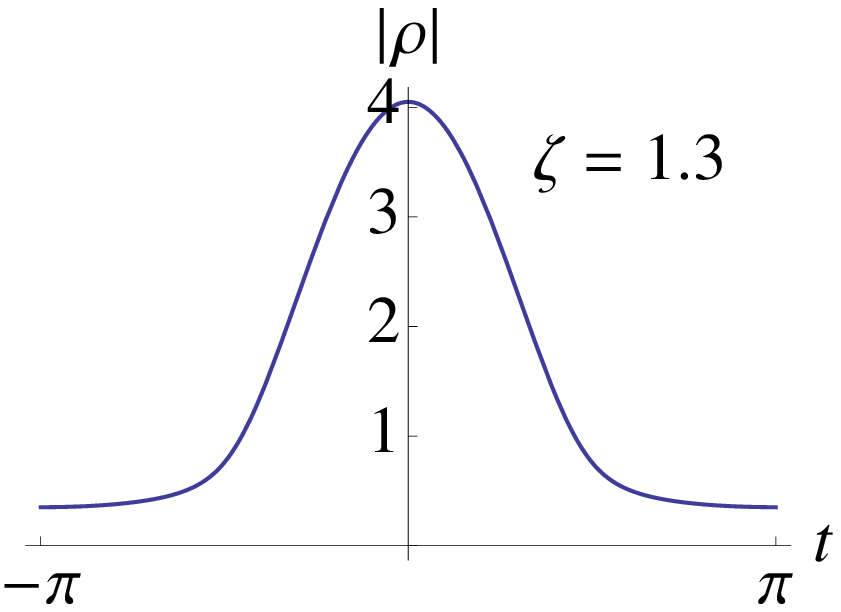, width =1.7in}
\hspace{0.2in}
\epsfig{file=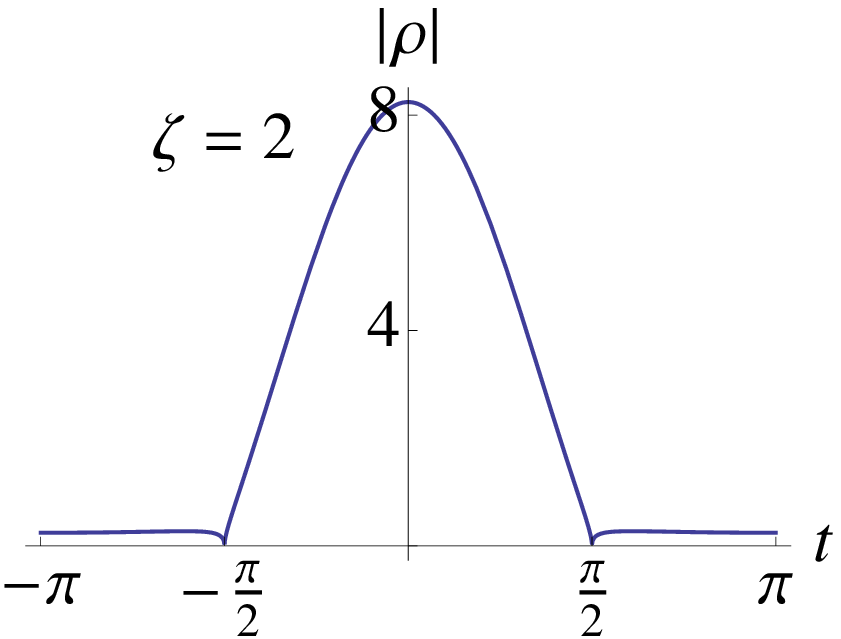, width =1.9in}
\hspace{0.2in}
\epsfig{file=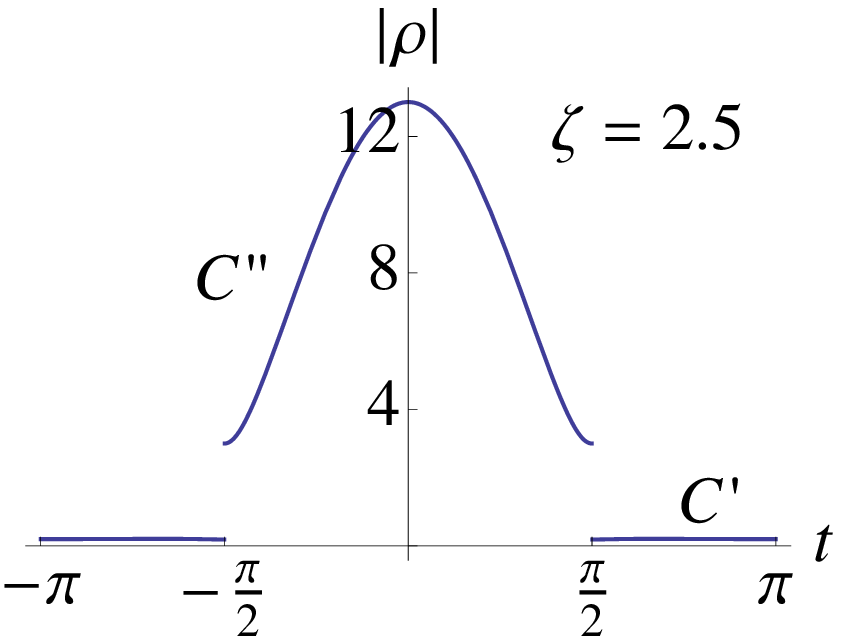, width =1.7in}
\end{center}
\caption{\small{The absolute value of $\rho(z(t))$ as a function of $t$ as the fugacity is increased, for $\nu_S= 0.5$. When $\zeta=2$, the density vanishes at $t=\pm \tfrac{\pi}{2}$, whilst for larger values, a discontinuity develops, between the (high) density along the smaller contour ${\cal C}''$ and the larger one. The magnitude of the density along ${\cal C}'$ remains small but always non-vanishing.}}
 \label{density} 
 \end{figure}

For large $\zeta$, Eq.\eqref{implicitc} has two qualitatively distinct solutions for $z(t)$,
\bea
z(t)\,&\simeq& \,\frac{1}{\zeta}\,\left(1-\frac{e^{i t/\nu _S}}{\zeta^{\nu_S}}\,\right)\,,\qquad\qquad -\pi\nu_S \leq t \leq \pi\nu_S\,,\\\nonumber\\\nonumber
&\simeq& \, \zeta^{\frac{\nu_S}{1-\nu_S}}\,\exp\left(i\,{\tfrac{t-{\rm sgn}(t)\pi\nu_S}{1-\nu_S} }\right)\,,\qquad \nu_S\pi < |t|\leq \pi\,.
\eea
The first of these  has $z\simeq \zeta^{-1} \ll 1$, whilst the second has $|z|\simeq \zeta^{\nu_S/1-\nu_S} \gg 1$. These are the two contours ${\cal C}''$ and ${\cal C}'$ in the limit of asymptotically large fugacities.  The $2 N_S$ eigenvalues making up the smaller contour ${\cal C}''$ have essentially collapsed to a point with parametrically small VEVs, whilst those lying on ${\cal C}'$ are effectively uniformly distributed on a large circle. 
Correspondingly, the magnitude of the eigenvalue density $|\rho(z)|$ exhibits a discontinuity for $\zeta > \zeta_{\rm crit}$ as shown in Figure \eqref{density}.

The physical interpretation of this non-smooth behaviour in the eigenvalue distribution is not immediately obvious. One possibility, which we have not explored, is that beyond $\zeta=\zeta_{\rm crit}$ a new saddle point solution with a gapped eigenvalue distribution appears, on disjoint open contours. For the moment we will simply follow our original saddle point configuration beyond $\zeta_{\rm crit}$, where it splits, and attempt to understand its physical interpretation.

Despite the singular behaviour of the eigenvalue distribution, gauge invariant quantities such as the baryon number defined via Eq.\eqref{number}, $\langle\Tr U^n\rangle$ and $\langle\Tr U^{\dagger n}\rangle$ remain unchanged as $\zeta$ is dialled past $\zeta_{\rm crit}$. Here we are specifically referring to multiply wound loops with $n$ fixed as $N$ is taken to infinity. The contributions to these quantities from the pole at $z=1/\zeta$, cancel between the two disjoint contours.
This is because ${\cal C}'$ and 
${\cal C}''$ must naturally be assigned opposite orientations when computing physical observables using the density function $\rho(z)$. 
In particular, the total baryon/squark number remains zero during this process. 
The physical reason for this can be traced to the fact that the only matter modes we have kept in our discussion are the squark zero modes. When 
$2 N_S < N$, no gauge-invariant state carrying baryon number can be made solely using the complex scalars and therefore baryon number cannot change in this regime. We will see later how this picture is altered by the inclusion of additional degrees of freedom  in the effective description.

While the total baryon/squark number is zero, the two components of the eigenvalue distribution each carry  equal but opposite baryon number:
\be
N\,{\cal Q}[{\cal C}''] = -\,N\, {\cal Q}[{\cal C}']=\,
N\,\nu_S(1-\nu_S)\,.
\ee

It is interesting to ask whether the splitting of the eigenvalue distribution into smaller ``bubbles'', can be characterized by means of gauge-invariant order parameters as a phase transition. Possibly, 
Polyakov loops in rank $k$ tensor representions, symmetric or antisymmetric, with $k \sim{\cal O}(N)$, could exhibit specific non-analyticities associated to the splitting of the distribution. Such observables are known to encode gapping transitions of the Gross-Witten-Wadia (GWW) type \cite{Paniak:1997yt}, and can be computed using  methods 
explicitly discussed in \cite{Hartnoll:2006is}. We postpone further study of these issues for future work.

Noting that $\zeta_{\rm crit}$ is strictly greater than unity for $\nu_S < 1$, a surprising feature of the saddle point configuration above, is that it does not exhibit any obvious non-analyticities when $\zeta=1$. When $\zeta>1$, the chemical potential $\mu$ exceeds the energy $\e_0$ of the lightest mode and, in a free theory, should destabilize the theory by inducing runaway VEVs for the squark zero modes. In particular the value of $\mu \simeq\e_0$ at which the splitting of the distributions occurs (at low temperatures),
\be
\mu = \e_0 + T \ln\zeta_{\rm crit}\,,
\ee 
is strictly greater than $\e_0$, and yet we did not encounter any sign of an instability as the $\mu$ was dialled past $\e_0$.

One possible reason why the putative instability is not visible in the finite volume theory, has to do with gauge invariance. The squark fields are not gauge-invariant and when $2 N_S < N$, it is not possible to form gauge-singlet baryon operators in the matrix model truncated to the zero modes. Therefore no gauge-invariant order parameter exists to signal the onset of an instability (e.g. a runaway VEV) in the truncated model. 

From the point of view of the effective matrix model for $U$, a squark VEV instability is evaded by non-zero VEVs for the Polyakov loop eigenvalues in the confined phase. In the free theory, the action for the scalar zero modes, in the presence of the chemical potential and a diagonal VEV for $A_0$ as in \eqref{eval}, has the form
\be
{\cal L}_{S}\,=\,\sum_{j=1}^{N_S} (\partial_\tau q^{\dagger}_j+i\,T\alpha\cdot q^\dagger_j +\mu q^\dagger_j)\left(\partial_\tau q_j - i\,T\alpha\cdot q_j -\mu q_j\right) + (m_S^2 + \Xi R^{-2})q^\dagger_j q_j + (q\to \tilde q)\,. 
\ee
While the chemical potential acts as a negative mass squared for the scalar zero modes, non-vanishing eigenvalues of the Polyakov loop ($\alpha \equiv \{\theta_a\}$) appear to want to stabilize the theory. 

\subsubsection{ Gross-Witten-Wadia transition for $\nu_S>1$} 
\begin{figure}[h]
\begin{center}
\epsfig{file=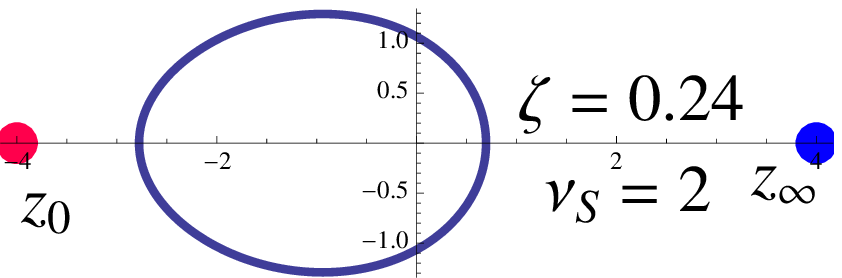, width =1.7in}
\hspace{0.3in}
\epsfig{file=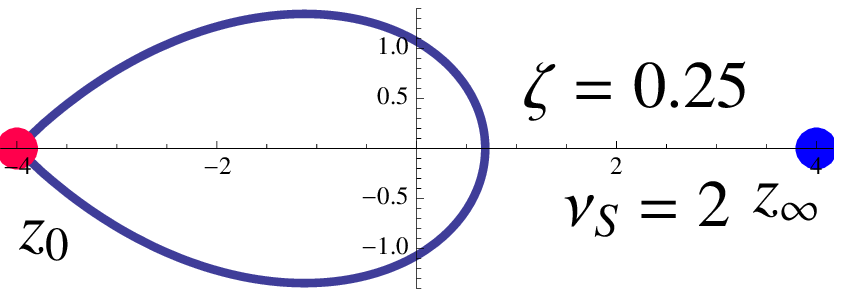, width =1.7in}
\hspace{0.3in}
\epsfig{file=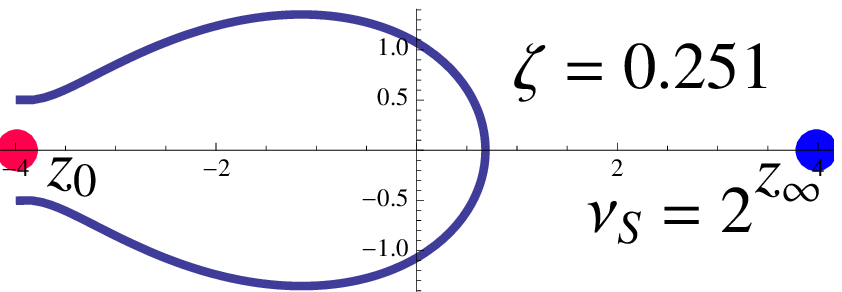, width =1.7in}
\end{center}
\caption{\small{For $\nu_S>1$ (number of squark flavours larger than the number of colors), the eigenvalue distribution develops a gap as $\zeta$ is increased past $\zeta_{\rm crit}$. The rightmost figure is just a continuation of the ungapped result beyond $\zeta_{\rm crit}$. The correct gapped distribution is determined separately below.}}
 \label{gw} 
 \end{figure}

When the number of charged scalar flavours $2N_S$ is larger than $N$, the number colors, we encounter a very different picture, with a new continuous deconfinement phase transition. 
Now the zero $Z_0$, of the analytic function $\rho(z)$, 
lies on the negative real axis. As $\zeta$ is increased, $Z_0$ moves toward, and eventually sits on the contour ${\cal C}$, precisely when the fugacity approaches the critical value
\be
\zeta_{\rm crit}\equiv \frac{(\nu_S -1)^{\nu_S-1}}{{ (\nu_S)}^{\nu_S}}\,  < 1\qquad {\rm for}\quad \nu_S>1\,.
\ee
The critical value follows from Eq.\eqref{implicitc}, upon substituting  in the expression for $Z_0$.
Figure \eqref{gw} illustrates this phenomenon for $\nu_S=2$. 
 At this point the eigenvalue distribution has a zero and wants to develop a gap. 
Finding the new distribution and establishing its existence requires a separate 
 analysis since the condition that ${\cal C}$ be closed, needs to  be relaxed. Before we calculate the gapped distribution for the new phase, we explain the physical origin of this transition. 

\subsubsection{Baryons ``melting'' at large-$\boldsymbol N$}
The putative non-analyticities we have seen so far, in the large-$N$ free theory in finite volume, can be provided a physical interpretation by looking at the grand-canonical partition function of  gauge-singlet states carrying baryon number. We first look at the case where the number of squark flavours is larger than the number of colours\footnote{The arguments presented here  grew out of a stimulating discussion initiated by S. Minwalla.}. In this case ($\nu_S > 1$), the theory has gauge-invariant excitations carrying baryon number,
\be
B_{S\,(i_1 i_2\ldots i_{N})}\,=\, q^{a_1}_{i_1}\,q^{a_2}_{i_2}\,\ldots 
\tilde q_{i_l}^{\dagger\,a_l}\ldots q^{a_N}_{i_N}\,\epsilon_{a_1 a_2 \dots a_N}\,.
\ee
The subscript `$S$' denotes that the baryon is made of scalars alone; in fact, from the zero modes of squark fields carrying baryon number $+1$. In our notation these are the $q_i$ and $\tilde q_i^\dagger$ with $i=1,2,\ldots N_S$. In the truncated low temperature model, 
when $2N_S > N$, the degeneracy of single baryon states is
\be
{\mathfrak D}(B_S)\,=\,\binom{2N_S}{N} \approx \exp\left[N\left(\nu_S\ln\nu_S-(\nu_S-1)\ln(\nu_S-1)\right)\right].
\ee
Therefore the degeneracy  of baryon-like states grows exponentially with $N$ in the Veneziano limit. 
Each of these states has mass $N\e_0 $ and $U(1)_B$ charge equal to $N$. Hence, the partition sum for  gauge-invariant states at low temperatures wherein the ``lightest states'' charged under $U(1)_B$ contribute goes as,  
\be
{\cal Z}\,=\,1+
\exp\left[N\left(\nu_S\ln\nu_S-(\nu_S-1)\ln(\nu_S-1)\right)\right]\times e^{-N\beta(\e_0-\mu)}+\ldots\,.
\ee
In the large $N$ limit, the partition function diverges exponentially with $N$, when
\be
\mu > \e_0 - T \left(\nu_S\ln\nu_S-(\nu_S-1)\ln(\nu_S-1)\right)
\ee
which coincides with the critical value of the fugacity 
$\zeta_{\rm}$ ($\nu_S>1 $) for the onset of a putative GWW transition in our matrix model. We will see below that this phase transition is similar to a deconfinement transition for squarks. At large $N$, due to the exponential growth of baryon operators, beyond the critical value of the chemical potential baryons are not the correct degrees of freedom, and the theory enters the phase of cold deconfined squarks.

When the number of squark flavours is smaller than the number of colours, it is not possible to form a baryon operator out of the squark fields (such operators vanish identically for $2N_S < N$). However it is still possible to form {\it adjoint-baryons}, if we include adjoint modes.  On the three-sphere, such exotic baryon-like states are possible even if the theory has no matter fields in the adjoint representation. The harmonics of the gauge field naturally provide such objects, and can be used to make adjoint-baryons.

In particular, using a field $\Phi$ transforming in the adjoint representation, we can make composite fields of the form 
$\Phi^n\,q_i$ and $\Phi^n\,\tilde q^\dagger_i$, transforming in the fundamental representation of $SU(N)$. It is easily seen that the lightest adjoint-baryon we can construct for $2N_S < N < 4 N_S$, is
\be
{B}_{{\rm Ad}\,(i_1 i_2\ldots)} = q_1^{a_1} q_2^{a_2}\ldots(\Phi q_{i_1})^{b_1}
(\Phi q_{i_2})^{b_2}\ldots \epsilon_{a_1 a_2\ldots a_{2N_S}b_1 b_2\ldots b_{N-2N_S}}\,.
\ee
For theories with $4N_S < N $ or $2\nu_S <1 $ we will also need to include states made from contractions of higher powers of the adjoint mode, such as $\Phi^2 q_i$, $\Phi^3q_i$, etc.
Taking the mass of the field $\Phi$ to be $\e$ (this could be a scalar zero mode or the lightest mode of the gauge field), 
the lightest of the adjoint-baryon operators has a mass
$N(\e_0 + (1-\nu_S)\e)$. The degeneracy of these adjoint baryons also grows exponentially with $N$ as\footnote{At this stage, for simplicity, we are assuming that the number of such light adjoint fields is one. If the number of adjoint species is $d$, the degeneracy of adjoint baryons grows as  $\binom{2N_S\,d}{N-2N_S}$},
\be
{\mathfrak D}(B_{\rm Ad})\,=\,\binom{2N_S}{N-2N_S}\approx
\exp\left[N\,\ln
\left(\frac{\nu_S^{\nu_S}}{(1-\nu_S)^{1-\nu_S}\,(2\nu_S-1)^{2\nu_S-1}}\right)\right]\,.
\ee
The exponential growth implies a phase transition when the chemical potential is increased beyond the critical value,
\be
{\mu}_{\rm crit}\,=\,\e_0+(1-\nu_S)\e- 
\frac{T}{N}\,\ln {\mathfrak D}(B_{\rm Ad})\,.
\label{adjoint}
\ee 
It is interesting to see how the onset of this transition and the behaviour of the theory beyond it, is captured by the matrix model description. 
Of course, this will require going beyond the truncated model for the ``lightest modes'' in the presence of the chemical potential. In fact if  the Kaluza-Klein harmonics of the squarks are lighter than all adjoint modes, similar phase transitions can be triggered by the exponential growth of baryon states made from squarks carrying different angular momentum ($\ell$) quantum numbers. 
We postpone further discussion of this to Section 4.

In order to better understand the nature of phase transitions induced by a chemical potential in this theory , we return to study the truncated matrix model in its gapped phase for $2N_S> N$.

\subsubsection{Gapped Phase}
\label{sec:truncatedgap}
Let us now look for the gapped solution to the saddle point equation, which we expect to find when $2 N_S>N$ and the fugacity exceeds the critical value beyond which the large-$N$ degeneracy of baryon operators leads to a divergent partition function. The contour ${\cal C}$ along which the eigenvalues are distributed, must be taken to be open. We begin by defining the resolvent function,
\be
\omega(z)\equiv - \frac{1}{N}\sum_{a=1}^n\frac{z+z_a}{z-z_a}\, \to \,- \int_{\cal C}\frac{dz'}{2\pi i}\,\rho(z')\,\frac{z+z'}{z-z'}\,.
\label{resolvdef}
\ee
The resolvent function will have a branch cut along the contour ${\cal C}$, and importantly, branch points corresponding to the end-points of ${\cal C}$. In particular, the saddle point equation, when expressed in terms of the resolvent, shows that the branch points must  be of the square root type:
\be
-\frac{1}{2}\left[\omega(z+\epsilon) +\omega(z-\epsilon)\right]\,=\,
{\cal Q}-\nu_S\,\zeta\,\frac{z}{1-\zeta\, z}\,,\qquad
z\in {\cal C}\,.
\ee
Here $z\pm\epsilon$ are two points infinitesimally close to the cut ${\cal C}$, on either side of it. Although, generally speaking, branch cuts of complex functions can be moved around, in the present case there is one special choice of ${\cal C}$ which is determined uniquely by \eqref{defrho} once the spectral density function $\rho(z)$ known. The density function is also given by the discontinuity of the resolvent across its branch cut:
\be
{\frac{1}{2}[\omega(z+\epsilon)-\omega(z-\epsilon)]\,=\,z\,\rho(z)}\,,\qquad z\in{\cal C}\,.
\ee
This is a very useful relation, as it allows us to re-express the expectation values of matrix model observables as contour integrals enclosing the branch cut along ${\cal C}$. In particular, line integrals along ${\cal C}$ involving the spectral density $\rho(z)$ can be replaced by contour integrals involving the resolvent function:
\be
\oint _{\hat{\cal C}} \frac{dz}{4\pi i}\, \omega(z)\,\frac{1}{z}\ldots\,=\,\int_{\cal C}\frac{dz}{2\pi i}\,\rho(z)\ldots\,,{\qquad}\hat{\cal C}\,\,{\rm enclosing}\,\, {\cal C}\,.
\label{resolvent}
\ee

From our analysis in the ungapped phase, we also expect the density function $\rho(z)$ to be symmetric under reflection about the real axis, so that $\rho(z)=\rho(z^*)$. Therefore $\omega(z)$ must be of the general form
\be
\omega(z)\,=\,f(z)\sqrt{(z-a)(z-a^*)}+\nu_S\zeta\,\frac{z}{1-\zeta\, z}-
{\cal Q}
\ee
where $f(z)$ is an analytic function, which is completely determined by the requirement that $\omega(z)$ be regular at $z=\frac{1}{\zeta}$ and the asymptotics at infinity and the origin, namely,
\be
\lim_{z\to 0}\omega(z)\,=\,1\,,\qquad\qquad\lim_{z\to\infty}\omega(z)\,=\,-1\,.
\ee
We find that
\bea
&&f(z)\,=\, \frac{\nu_S}{(z\zeta-1)|a-\zeta^{-1}|}\,,
\label{extent}\\\nonumber
\\\nonumber
&&a \,=\, \frac{{\cal Q}^2+{\cal Q}\,\nu_S-\nu_S+1}{\zeta({\cal Q}+\nu_S-1)^2}\,+\,2i\,\frac{\sqrt{{\cal Q}({\cal Q}+\nu_S)(\nu_S-1)}}{\zeta({\cal Q}+\nu_S-1)^2}\,.
\eea
The final ingredient in the solution is the baryon number ${\cal Q}$ in the gapped phase. This is fixed by demanding the unit determinant condition on the matrix $U$, after expressing Eq.\eqref{norm} as an integral over a contour $\hat {\cal C}$ enclosing ${\cal C}$,
\be
\oint_{\hat {\cal C}} \frac{dz}{4\pi i}\, \ln z\, \frac{\omega(z)}{z}\,=\,0.
\ee
Since $\omega(z)$ itself has no singularities other than the branch cut enclosed by $\hat{\cal C}$, the latter contour can be deformed to enclose the other singularities of the integrand -- the poles at $z=0$ and $z=\infty$, and the branch cut of $\ln z$ which must be taken to run along the negative real axis.
Careful evaluation of the corresponding contour integrals 
yields{\footnote{The relative position of ${\cal C}$ with respect to the branch cut of $\ln z$, plays an important role in obtaining the correct result.}${}^,$\footnote{This relation between $\zeta$ and $Q$ suggests an underlying combinatorial origin, from Stirling's approximation, 
$n! \sim n^n$.
}}
\be
\boxed{\zeta\,=\,\frac{({\cal Q}+\nu_S-1)^{{\cal Q}+\nu_S-1}}{({\cal Q}+\nu_S)^{{\cal Q}+\nu_S}}\,\frac{({\cal Q}+1)^{{\cal Q}+1}}{{\cal Q}^{\cal Q}}\,.}
\label{bnumber}
\ee

\begin{figure}[h]
\begin{center}
\epsfig{file=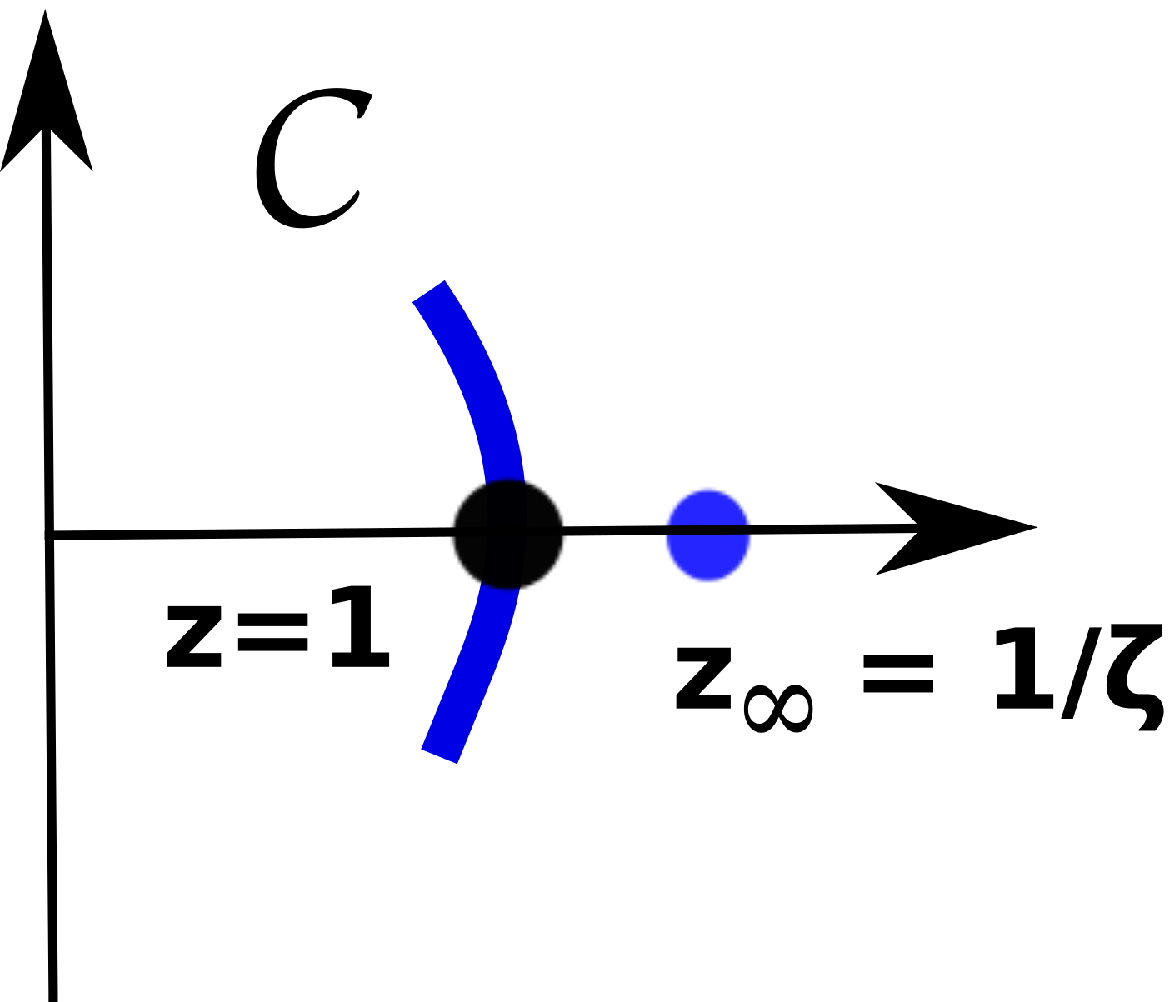, width =1.5in}
\hspace{0.3in}
\epsfig{file=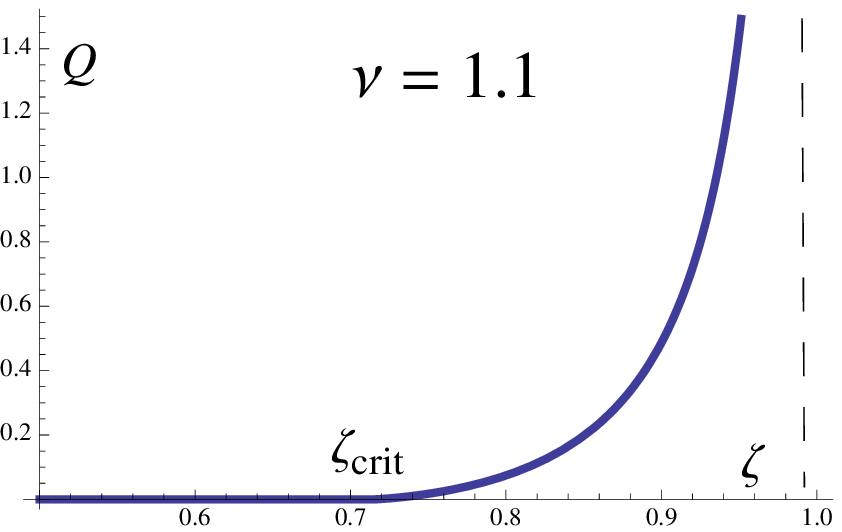, width =1.7in}
\hspace{0.3in}
\epsfig{file=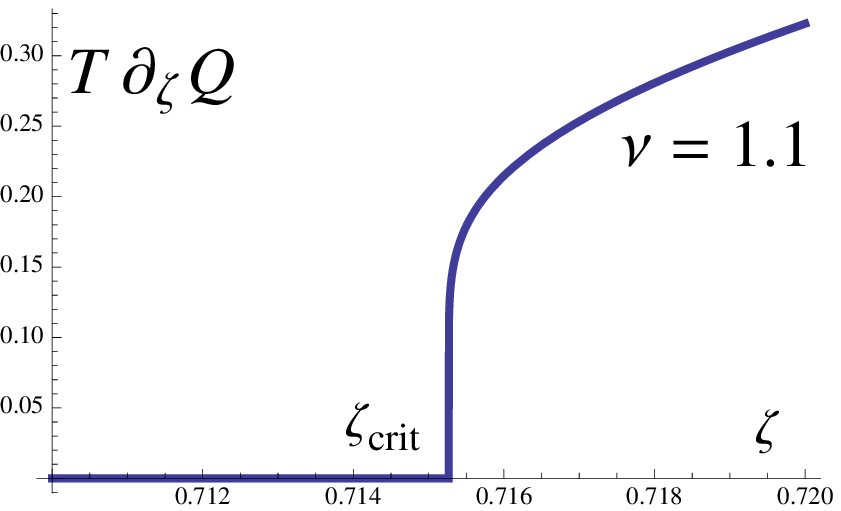, width =1.7in}
\end{center}
\caption{\small{ \underline{\bf Left}: A depiction of the gapped distribution as the fugacity $\zeta$ is increased towards $1$. It shrinks to the point $z=1$ in the limit $\zeta\to 1$, when the pole $Z_\infty$ collides with the eigenvalue distribution. \underline{\bf Centre}: The baryon number becomes non-zero when $\zeta > \zeta_{\rm crit}$ and eventually diverges at $\zeta=1$, signalling the onset of Bose condensation. \underline{\bf Right}: Continuity of second derivative of the grand potential (or first derivative of ${\cal Q}$) across the third order transition. }} 
 \label{bosecon} 
 \end{figure}

The transition to the gapped phase occuring at $\zeta=\zeta_{\rm crit}$ is a third order phase transition, similar to the GWW model. The derivative of ${\cal Q}$ with respect to $\mu$, which is the second derivative of the grand potential, is continuous across the gapping transition,
\be
T\,\frac{\partial {\cal Q}}{\partial\mu}\,=\,
\frac{T^2}{N^2}\,\frac{\partial^2\ln{\cal Z}}{\partial\mu^2}\,=\,\frac{1}{-\ln {\cal Q} +\ln({\cal Q}+1)+\ln\left(\frac{{\cal Q}+\nu_S-1}{{\cal Q}+\nu_S}\right)}.
\ee
At the transition point where $\zeta=\zeta_{\rm crit}$ and ${\cal Q}$ is vanishing, the second derivative of the grand potential also vanishes and is continuous, while the third derivative is discontinuous and diverges when $\zeta>\zeta_{\rm crit}$ (see Figure\eqref{bosecon}).

We can now see in Figure\eqref{bosecon} what happens when the fugacity $\zeta$ is increased beyond the critical value $\zeta_{\rm crit}$ towards $\zeta =1$. As $\zeta$ approaches unity (equivalently the chemical potential approaches the energy $\e_0$ of the lightest mode), the baryon number ${\cal Q}$ diverges. This is the onset of {\em Bose condensation} of squarks.  Furthermore as ${\cal Q}\to \infty$, the contour ${\cal C}$ shrinks to a point because the branch points at $z=a$ and $z=a^*$, both approach the point $z=1$ on the real axis.

 The third order transition we have seen above has a natural interpretation as deconfinement of squarks or fractionation of baryons. The ``holomorphic'' observables ${\rm Tr}\, U^n$ which vanished in the ungapped phase, are now non-zero. 
 Equations \eqref{polloops1} and \eqref{resolvent} imply that expectation values of these are encoded in the Laurent series expansion of $\omega(z)$ at infinity,
 \be
 \omega(z)\,=\,-1 - 2\, \sum_{n=1}^\infty 
 \langle\tfrac{1}{N}\Tr U^n\rangle\,\frac{1}{z^n}\,. 
\ee
For example,
\be
\langle\tfrac{1}{N}\Tr U\rangle\,=\,
\frac{1}{\zeta}\frac{\cal Q}{{\cal Q}+\nu_S-1}\,,
\ee
which approaches unity as $\zeta \to 1$ and $Q\to \infty$. 
Since the chemical potential breaks the symmetry under baryon charge conjugation, the charge conjugate set of observables ${\rm Tr}\, U^{\dagger n}$, should have different expectation values. Once again, Equations \eqref{polloops1} and \eqref{resolvent} fix them in terms of the Taylor series expansion of $\omega(z)$ around the origin:
\be
\omega(z)\,=\, 1 + 2 \sum_{n=1}^\infty\,\langle\tfrac{1}{N}\Tr U^{\dagger n}\rangle\, z^n \,,
\ee
so that $\langle\tfrac{1}{N}\Tr U^\dagger\rangle= \zeta\,\frac{{\cal Q}+\nu_S}{{\cal Q}+1}$, is distinct from its holomorphic counterpart. 
Since all the Polyakov loops (singly and multiply wound), have non-vanishing expectation values, the gapped distribution describes a deconfined phase of squarks \cite{svetitsky}.

It is interesting to note that in the limit $\zeta\to 1$, at the onset of the Bose condensed phase, the branch cut 
${\cal C}$ shrinks to the point $z=1$, and the resolvent simplifies,
\be
\lim_{\zeta \to 1}\omega(z)\,=\,\frac{1+z}{1-z}\,.
\ee
In this limit all single trace expectation values become equal, $\langle\tfrac{1}{N}\Tr U^n\rangle=\langle\tfrac{1}{N}\Tr U^{\dagger n}\rangle=1$. 
This means that the angular variables $\{\theta_a\}$ all vanish and the gauge group is ``unbroken''. When the $\{\theta_a\}$ are distinct and non-zero, the gauge symmetry is broken to $U(1)^{N-1}$. 
\section{Incorporating heavy modes on $S^3$}

In the discussion above we saw the appearance of a third order deconfinement transition in the general situation where squarks (scalars in the fundamental representation) are the lightest modes. The non-analyticity of the grand potential as a function of baryonic chemical potential was demonstrated when the number of squark flavours ($2 N_S$) was larger than the number of colors ($N$). In this situation the non-analytic behaviour of the grand canonical partition function is attributed to the exponentially large degeneracy ($\sim e^N$) of baryon operators.

When the number of squark flavours is small, i.e. $2 N_S < N$, the situation is somewhat subtle since baryon operators made from squark zero modes alone, do not exist. In this case, we have already seen that the partition function truncated to squark zero modes (the ``lightest'' degrees of freedom with $U(1)_B$ charge $+1$) 
displays no non-analyticities as the chemical potential is increased (despite  ``bubble splitting'' of the eigenvalue distribution). However, baryon-like states, namely adjoint-baryons, do exist when then theory has additional adjoint modes which are neutral under $U(1)_B$. In the Veneziano limit, the number of adjoint-baryons of a given mass also grows exponentially with $N$, possibly leading to a phase transition \eqref{adjoint}. Below we will demonstrate, using the approach of holomorphic matrix integrals, the existence of deconfinement transitions also when $2 N_S < N$, when additional modes on $S^3$ are included in the description. 
 For sufficiently light squarks, another effect can become important: it is possible to form baryons using squarks with differing $\ell$ quantum numbers. These also exhibit an exponentially large degeneracy, and should be responsible for transitions in the large-$N$ matrix model.

\subsection{The effect of adjoint modes}
\label{sec:adjoints}
We now explore what happens when light adjoint bosonic fields {\em neutral under $U(1)_B$}, are included in the effective theory for $U$. At the outset, we emphasize that such adjoint modes will appear in all gauge theories - they can be viewed either as arising from  harmonics of free vector fields on $S^3$ or from additional adjoint matter fields (as in ${\cal N}=2$ supersymmetric theories).
Including their lowest harmonics in our effective description at sufficiently low temperatures, the effective action for the unitary matrix $U$ is
\bea
&&S_{\rm eff}\,=\, N\,\nu_S\,\Tr\ln(1-\zeta U) -d\sum_{n=1}^\infty\frac{x^n}{n}\,{\rm Tr}\,\, U^n{\rm Tr}\,\, U^{\dagger n}\,
\label{action2}
\\\nonumber
&& x\,=\,e^{-\e/T}\,,\qquad\zeta\,=\,e^{(\mu-\e_o)/T}
\eea
where $d$ is the number of such adjoint species and $\e$ is the associated energy. Here, we are implicitly assuming that 
there is always a range of parameters for which the anti-squarks, and other harmonics of fields on $S^3$ can be consistently ignored. We will eventually determine values of the chemical potential for which the modes retained above are the most important ones.

The fugacity is defined as before, always with reference to the energy $\e_0$ of the lowest squark mode charged under $U(1)_B$.  For the lowest harmonics of vector fields on $S^3$, $d=6$ and $\e= 2/R$, while for conformally coupled massless scalars, $\e=1/R$ (see Table\eqref{table}). 
In what follows, for simplicity we set $d=1$ (one adjoint species). 
At this juncture we impose no special restrictions on the chemical potential or the fugacity.
Notice that 
in the low temperature limit, $x$ is exponentially small and the adjoint contributions appear to be exponentially suppressed. Despite this, we will see that when the fugacity $\zeta$ is cranked up to sufficiently large values, the adjoint modes drive a GWW deconfinement phase transition. 


The large-$N$ equation of motion for the eigenvalues of $U$, following from the modified effective action \eqref{action2}, can be written in holomorphic form as before,
\bea
&&\oint_{\cal C}\frac{dz^\prime}{2\pi i}\,
\rho(z')\left[{\cal P}\frac{z+z'}{z-z'}+x\left(\frac{z}{z'-x\,z}-\frac{z'}{z-x\,z'}\right)\right] + 
\frac{\nu_S\,\zeta\,z}{1-\zeta z }\,=\,{\cal Q}\,,
\\\nonumber
\\\nonumber
&&z\in{\cal C}\,.
\eea
At low temperatures $x\ll 1$ and $\zeta\ll1$, the contour 
${\cal C}$ is closed and all the poles in $\rho(z)$ can be safely taken to lie outside ${\cal C}$. Interestingly, this equation can also be solved exactly to yield $\rho(z)$:
\be
\boxed{
\rho(z)\,=\,\frac{1}{z} + \zeta\nu_S\,\sum_{n=0}^\infty
\frac {x^n}{1-\zeta\,z\,x^n}\,,\qquad {\cal Q}=0\,}.
\label{exactrho}
\ee 
The infinite sum over $n$ can be rewritten in terms of the q-Pochammer symbol as $\tfrac{d}{dz}\,(\zeta z\,;\,x)_\infty$. The contour ${\cal C}$ itself is then given by $z(t)$, where, using Eq.\eqref{defrho}
\be
e^{it}\,=\, \frac {z}{\prod_{n=0}^\infty\left(1-\zeta\,z\,x^n\right)^{\nu_S}}\,=\,\frac{z}{\left[\left(\zeta\,z\,;\,x\right)_\infty\right]^{\nu_S}}
\label{exactc}
\ee

To understand the physics of this system systematically, working consistently at low temperatures $(x\ll 1)$, let us begin by keeping only the first two terms $(n=0,1)$ of the infinite sum,
\be
\rho(z)\,\simeq\,\frac{1}{z} + \zeta\nu_S\,\left(\frac {1}{1-\zeta\,z }+
\frac {x}{1-\zeta\,z\,x}+\ldots\right)\,.
\ee
In the language of counting baryon-like states, this means we only 
count adjoint-baryons made from $2N_S$ squarks and $(N-2 N_S)$ ``adjoint-squarks'' of the form $\Phi\cdot q_i$, each with a single adjoint matrix.
  
Clearly, this goes one step beyond the truncated model of Section 3. As before, we can first locate the poles and zeros of this function. For small enough $\zeta$ and $x$, the two simple poles at $Z_\infty^{(1)}={\zeta}^{-1}$ and $Z_\infty^{(2)}=(x\,\zeta)^{-1}$ are far from the (closed) contour ${\cal C}$ which encloses the origin. Our approximate form of $\rho(z)$ also contains two zeroes, which, for $x \ll 1$, are at 
\be
Z_0^{(1)}\simeq\frac{1}{\zeta(1-\nu_S)}\,,\qquad \quad 
Z_0^{(2)} \simeq \frac{1}{x}\,\frac{1-\nu_S}{\zeta(1-2\nu_S)}\,,
\qquad x\ll1\,.
\ee
The first zero lies to the right of ${\cal C}$ when $2N_S < N$ and we have already observed its effect in the context of the splitting of the eigenvalue distribution. We denote the critical value of $\zeta$ at which the splitting of eigenvalues occurs as $\zeta_{\rm crit}^{(1)}$. At this value, the zero $Z_0^{(1)}$ hits the contour ${\cal C}$.

\begin{figure}[h]
\begin{center}
\epsfig{file=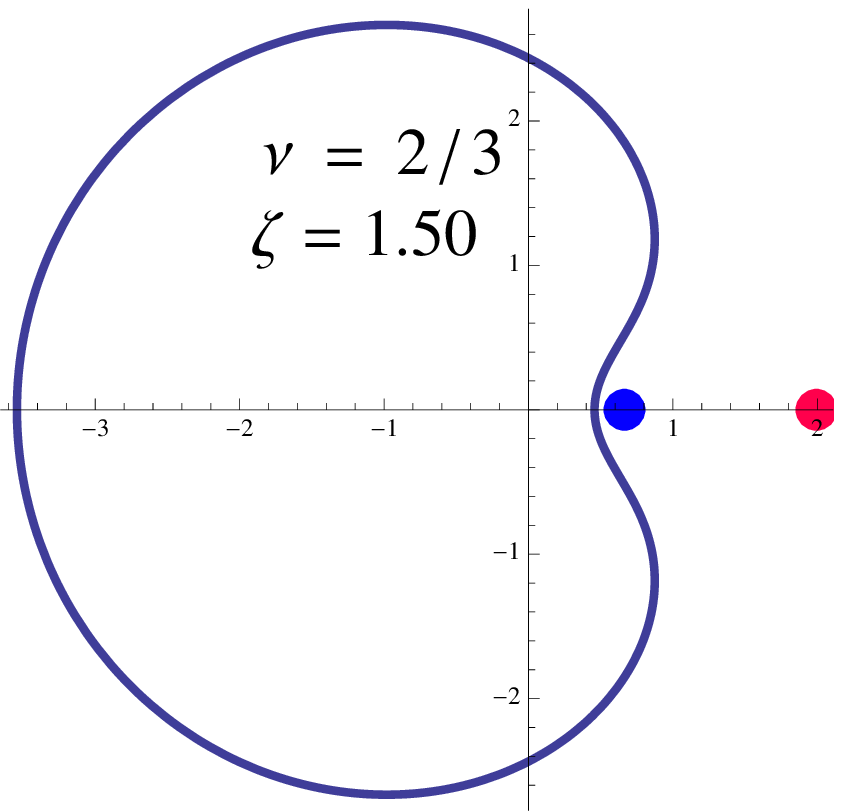, width =1.3in}
\hspace{0.4in}
\epsfig{file=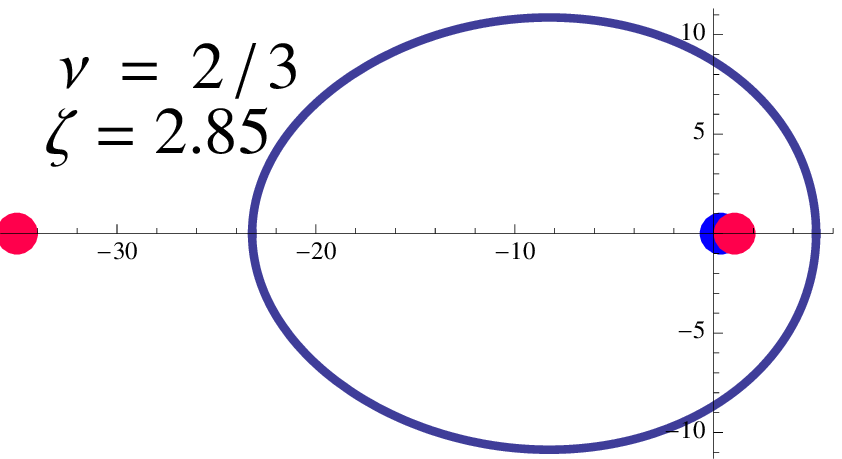, width =1.5in}
\hspace{0.4in}
\epsfig{file=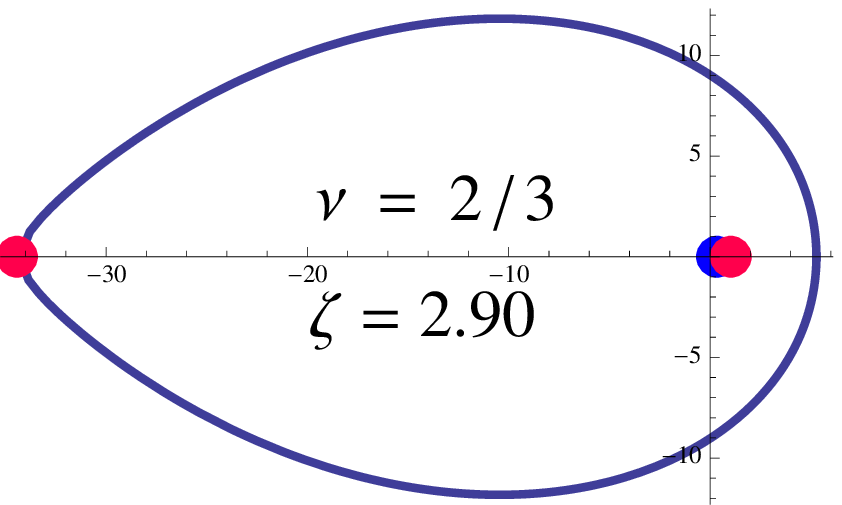, width =1.5in}
\end{center}
\caption{ Plots demonstrating the behaviour of the contour ${\cal C}$ of eigenvalues in the $z$-plane, with increasing fugacity $\zeta$, and the relative postions of poles (in blue) and zeroes (in red) of $\rho(z)$. At a critical value of $\zeta$, after the ``splitting'' of the contour, the distribution develops a zero, signalling the onset of a gap. The curves were obtained assuming $x=0.1$.
}
\label{adjointgap}
\end{figure}

The second zero is new and has appeared solely due to the inclusion of the adjoint modes in our description.
It is on the negative axis and to the left of ${\cal C}$ if $2N_S < N < 4N_S$; else it lies on the positive real axis. 
Let us focus attention on the case with $2N_S < N < 4 N_S$. 
As $\zeta$ is increased, $Z_0^{(2)}$ moves towards the contour ${\cal C}$ and eventually collides with it when (using Eq.\eqref{exactc})
\be
\zeta\, =\,\zeta_{\rm crit}^{(2)}\,\simeq\,
x^{\nu_S-1} \,(2\nu_S-1)^{2\nu_S-1} \,(1-\nu_S)^{1-\nu_S}\,
(\nu_S)^{-\nu_S}\,.
\ee
The corresponding value of the critical chemical potential,
\bea
&&\mu_{\rm crit}\,=\\\nonumber
&&\e_0+ (1-\nu_S) \e \,-
T\left(\nu_S\ln(\nu_S)-(2\nu_S-1)\ln(2\nu_S-1)-(1-\nu_S)\ln(1-\nu_S)\right)\,,
\eea
 coincides precisely with the value deduced from the exponential growth of the degeneracy of  adjoint-baryon states (Eq.\eqref{adjoint}) for the case $2N_S < N < 4 N_S$. 
 The phenomenon is summarized in Figure \eqref{adjointgap}

The preceding analysis demonstrates that even when the number of squark flavours is smaller than the number of colors, the theory on $S^3$ is subject to a low temperature GWW deconfinement transition when the chemical potential approaches a threshold value determined by the mass of the lightest adjoint-baryons and their degeneracy in the Veneziano limit. We also deduce that the effective action retaining only squark zero modes and the adjoints is a good description at low temperatures when $N\mu$ is in the vicinity of the mass of the lightest adjoint-baryon.

\subsection{Effect of higher squark harmonics}

When $2N_S < N$ it is obvious that baryons cannot be made from squark zero modes. However one may still use squark modes with differing angular momentum quantum numbers to form gauge-singlet baryons. Such baryons could be formed by $2 N_S$ squark zero modes $(\ell=0)$ and $N-2 N_S$ squark modes with $\ell=1$. Given a degeneracy $d$ for the $\ell=1$ scalar spherical harmonics,
the number of such baryons is $\binom{2N_S\,d}{N-2N_S}$ and much in the same way as for adjoint baryons, it should lead to a phase transition with increasing fugacity.

The baryons made from higher harmonics of squark modes become relevant only if the energies of the squark modes are significantly smaller than $R^{-1}$ so that all adjoint harmonics (including the gauge fields) are heavier. In this situation the effective action for $U$, keeping only the $\ell =0$ and $\ell=1$ squark modes, is
\bea
&&S_{\rm eff}\,=\,N\, \nu_S\,\Tr\,\ln(1-\zeta U) + d\, N\,\nu_S\,\Tr\,\ln(1-x\,\zeta U)\\\nonumber\\\nonumber
&& \zeta=e^{-\beta(\e_0-\mu)}\,,\quad x= e^{-\beta(\e-\e_0)}\,.
\eea
Here $\e$ is the energy of the $\ell=1$ harmonic, and the 
degeneracy of this mode $d=4$. Now the saddle point configuration in the ungapped, confined phase, is solved by the density
\be
\rho(z)\,=\, \frac{1}{z}+\frac{\zeta\,\nu_S }{1-\zeta\,z}+\frac{d\, x\,\zeta\,\nu_S }{1-x\,\zeta\,z}\,.
\ee
The density function has zeroes at 
\be
Z_0^{(1)}\simeq\frac{1}{\zeta(1-\nu_S)}\,,\qquad \quad 
Z_0^{(2)} \simeq \frac{1}{x}\,\frac{1-\nu_S}{\zeta(1- (d+1)\nu_S)}\,,
\label{zeroes}
\qquad x\ll1\,.
\ee
In the range $2N_S < N < (d+1) 2N_S$, the first zero lies on the positive real axis, whilst the second zero is on the negative real axis, to the left of the eigenvalue distribution. With increasing chemical potential, a zero in the eigenvalue distribution first appears when,
\be
\zeta\,=\,\zeta_{\rm crit}\,=\,x^{\nu_S-1}\,\frac{(1-\nu_S)^{1-\nu_S}((d+1)\nu_S-1)^{(d+1)\nu_S-1}}{(d\nu_S)^{d\nu_S}}\,,
\label{squarkcrit}
\ee
or
\be
\mu_{\rm crit}\,=\, \nu_S\,\e_0+(1-\nu_S)\,\e - T\,\ln\left(\frac{(d\nu_S)^{d\nu_S}}{(1-\nu_S)^{1-\nu_S}((d+1)\nu_S-1)^{(d+1)\nu_S-1}}\right)\,.
\ee
The critical value of the chemical potential is determined by the mass and degeneracy of the lightest baryon operators composed of the squark modes with $\ell=0$ and $\ell=1$. Below we will work out what happens to the system as the chemical potential is dialled past this critical value. In particular we will show the emergence of a gapped ``deconfined'' phase.

\subsection{Gapped Phase}
The transition to the gapped phase for $\nu_S < 1$ (small number of flavours) is always preceded by an eigenvalue splitting transition. The eigenvalues which have split off and formed the smaller ``bubbles'' do not appear to play any role in the subsequent dynamics as the chemical potential is increased to large values.

\subsubsection{Gapping/Bose condensation via higher harmonics}

Let us attempt to find the gapped distribution which results when the chemical potential is dialled past the critical value 
in Eq.\eqref{squarkcrit}. Since the distribution has already split prior to the gapping transition, we write the spectral density function as a sum of two (disjoint) pieces,
\bea
\rho(z)\,&=&\,\rho_1(z)\,,\qquad z\in {\cal C}'\\\nonumber 
&=&\,\rho_2(z)\,,\qquad z\in{\cal C}''\,.
\eea
where the first term is associated to the larger contour ${\cal C}'$, and the second term yields the density of the smaller ``bubble'' ${\cal C}''$ (see e.g. Fig.\eqref{contour}). In the splitting transition, a fraction $\nu_S$ of the $N$ eigenvalues break off to form ${\cal C}''$. We must therefore have
\be 
\int_{{\cal C}'} \rho_1(z)\,\frac{dz}{2\pi i}\,=\,1-\nu_S\,,\qquad
\oint_{{\cal C}''}\rho_2(z)\,\frac{dz}{2\pi i}\,=\,\nu_S\,.
\ee
From our earlier analysis, we know that the eigenvalue bubble is parametrically small, centred around $z\simeq \zeta^{-1}$, for $\zeta\gg 1$. The configuration at the large-$N$ saddle point solves
\be
{\cal P}\int_{{{\cal C}'}\cup \,{{\cal C}''}}\frac{dz'}{2\pi i}\,\rho(z')\,\frac{z+z'}{z-z'}\,+\,\frac{\nu_S\,\zeta\, z}{1-\zeta\, z}\,+\,d\frac{\nu_S\, x\,\zeta\, z}{1-x\,\zeta\,z}\,=\,{\cal Q}\,.
\ee
We further recall from the discussion on the eigenvalue splitting transition that the contour ${\cal C}''$ must be traversed in the clockwise sense.
In the low temperature limit, $x\ll 1$, with $\zeta$ large
\footnote{Note that $\zeta = \exp(-\beta(\e_0-\mu))$ where $\e_0$ is the energy of the $\ell=0$ mode. To see the gapping phase transition $\mu$ needs to be close to the mass of the baryon made from $\ell=0$ and $\ell=1$ harmonics. Due to the finite energy gap between the  $\ell=0$ and $\ell=1$ modes, $\zeta$ will naturally be parametrically large for low enough temperatures.}, the spectral densities in the two disjoint contours can be determined as follows. For $z\in {\cal C}'$, due to the small radius of ${\cal C}''$ ($\sim \zeta^{-1-\nu_S}$), the saddle point equation reduces to,
\be
{\cal P}\int_{{\cal C}'}\frac{dz'}{2\pi i}\, \rho_1(z')\,
\frac{z+z'}{z-z'}\,+\,d\frac{\nu_S \,x\,\zeta\,z }{1-x\,\zeta\,z}\,=\,{\cal Q}\,
\ee
where we have used the normalization condition that $\oint_{{\cal C}''} \rho_2(z)\frac{dz}{2\pi i}=\nu_S$. Remarkably, this is exactly the equation we solved to find the gapped phase of the truncated model in Section \ref{sec:truncatedgap}, provided we make the following replacements:
\be
\nu_S \to \frac{d\,\nu_S}{1-\nu_S}\,\qquad {\cal Q}\to 
\frac{{\cal Q}}{1-\nu_S}\,\qquad \zeta\to x\,\zeta\,.
\label{replacement}
\ee 
The factor of $(1-\nu_S)$ in the denominator accounts for the fact that $\rho_1(z)$ is a density for $(1-\nu_S) N$ eigenvalues making up ${\cal C}'$. The resolvent function for the gapped distribution $\rho_1(z)$ is
\bea
&&\omega_1(z)\,\equiv\, - \frac{1}{1-\nu_S}\,\int_{{\cal C}'}\frac{dz'}{2\pi i}\,\rho_1(z')\,\frac{z+z'}{z-z'}\,,\\\nonumber\\\nonumber
&&\omega_1(z)\,=\,f_1(z)\sqrt{(z-a_1^*)(z-a_1)} -\frac{{\cal Q}}{1-\nu_S} + \frac{d\,\nu_S}{1-\nu_S}\frac {x\, \zeta\, z}{1-x\,\zeta\, z}\,.
\eea
The function $f_1(z)$ and the positions of the branch points at $a_1$ and $a_1^*$ are given by Eq.\eqref{extent} with the replacements
indicated in Eq.\eqref{replacement}. It is useful to verify that right after the  onset of the gapping transition where  ${\cal Q}\to 0$, the two branch-points coalesce and 
\be
\lim_{{\cal Q}\to 0}\, a_1 = \frac{1}{x\,\zeta}\,\frac{1-\nu_S}{1-(d+1)\nu_S}\,,
\ee
 coinciding with the position of the zero (see Eq.\eqref{zeroes}) of the ungapped distribution at the onset of the phase transition. Therefore the solution we have found is indeed connected, across a continuous transition, to the ungapped/confined phase. 
 
 To find the distribution $\rho_2$, we pick a point $z\in {\cal C}''$ and noting that on this contour $z\sim \zeta^{-1}$, the saddle point equation becomes
 \be
 \oint_{{\cal C}''}\frac{dz'}{2\pi i}\,\rho_2(z')\,\frac{z+z'}{z-z'}\,- (1 -\nu_S) + \frac{\nu_S\,\zeta\,z}{1-\zeta\, z}\,=\,0\,.
 \ee
 Here we have used 
 $\int_{{\cal C}'}\rho_1(z)\frac{dz}{2\pi i}= (1-\nu_S)$. This equation is actually solved by the same distribution function we  found in the ungapped phase of the model truncated to zero modes,
 \be
 \rho_2(z)\,=\,\frac{1}{z}+\frac{\nu_S\, \zeta}{1-\zeta\, z}\,.
 \ee
 
 A final, important consistency check of our picture follows from the relation between the baryon number (${\cal Q}$) and the fugacity $\zeta$, in this phase. As in the basic truncated setup, the relation can be derived by enforcing the unit determinant condition on $U$,
 \be
 \int_{{\cal C}'\cup\,{\cal C}''}\frac{dz}{2\pi i}\,\rho(z)\,\ln\,z
 \,=\,0,
 \ee
and we find
\be
\boxed{
\zeta\,x^{1-\nu_S}\,=\,\frac{({\cal Q}-\nu_S+1)^{{\cal Q}-\nu_S+1}}{{\cal Q}^{\cal Q}}
\,\frac{({\cal Q}+\nu_S(d+1)-1)^{{\cal Q}+\nu_S(d+1)-1}}{{({\cal Q}+d\nu_S)}^{{\cal Q}+d\nu_S}}\,.}
\label{grandq}
\ee
This means that the baryon number diverges ${\cal Q}\to\infty$ when 
$\zeta\, x^{1-\nu_S}=1$ or $\mu\, = \,\nu_S \,\e_0+(1-\nu_S)\,\e$. Therefore when the baryon number chemical potential $N\mu$ equals the mass of the lightest baryon state that can be formed from the spherical harmonics of the squark modes ($\ell=0$ and $\ell=1$ in this case), Bose condensation of squarks occurs. We believe that it is the squarks which condense, because slightly before this happens (at low non-zero temperatures), the system  enters a deconfined phase by passing through a third order GWW transition. 

In all the GWW transitions we have encountered, the baryon number $N{\cal Q}$ is zero below the transition and begins to increase smoothly above it. Therefore, using ${\cal Q}=\tfrac{T}{N^2}
\tfrac{\partial\ln {\cal Z}}{\partial\mu}$, the grand potential is a constant independent of $\mu$ below the critical value of the chemical potential, and becomes a non-trivial function of $\mu$ above it. It follows from our results, such as, Eq.\eqref{grandq} that the grand potential above the transition is a complicated function of the form $-\beta \ln {\mathcal Z} = N^2 \,F(\nu_S, \mu,T)$. 
In contrast, in the low $\mu$ confined phase, we will see below that the grand potential is of the form $\sim N_S^2\,\tilde F(T,\mu)$, indicating that the contributions are from a gas of gauge singlet mesonic states.

\subsubsection{Gapping due to adjoint modes}

We have already shown that when the number of squark flavours is smaller than the number of colours, it is possible to make gauge-singlet adjoint baryons. The exponentially large degeneracy of these states drives the system towards a GWW ``gapping'' transition (see Section \ref{sec:adjoints}). Below we sketch the steps involved in finding the gapped phase of the model when light adjoint modes are included in the effective description. Since we expect a finite energy gap between the adjoint states and the squark zero modes, the GWW transition is preceded by the breaking up of the eigenvalue distribution into disjoint contours ${\cal C}'$ and ${\cal C}''$. As before we write $\rho_1(z)$, the gapped spectral density along ${\cal C}'$, in terms of a resolvent,
\be
\int_{{\cal C}'}\frac{dz}{2\pi i}\,\rho_1(z)\,=\,1-\nu_S\,,\qquad
\omega_1(z)\,=\,-\frac{1}{1-\nu_S}\,\int_{{\cal C}'}\frac{dz'}{2\pi i}\,\rho(z')\,\frac{z+z'}{z-z'}\,.
\ee
The resolvent, by definition, is analytic on the $z$-plane with its only singularity being a branch cut along ${\cal C}'$. The saddle point condition for the effective model with adjoint degrees of freedom, can then be rewritten as
\bea
&&-\frac{1}{2}\left(\omega_1(z+\epsilon)+\omega_1(z-\epsilon)\right) +
\frac{1}{2}\,\omega_1(x\,z) + \frac{1}{2}\,\omega_1(x^{-1}\, z)\,=\,\frac{{\cal Q} +\nu_S}{1-\nu_S}\,,\\\nonumber
&&z\in{\cal C}'\,,\qquad x\,=\, e^{-\beta\e}\,.
\eea
We have used the fact that $\frac{\zeta\,\nu_S\,z}{1-\zeta\, z}\simeq -\nu_S$ for large $\zeta$, and for $z\in {\cal C}'$. This is an interesting condition as it can be recast in a different form by defining a new analytic function,
\be
G(z)\,\equiv\,\omega_1\left(\sqrt x\,z\right) -\omega_1\left(\tfrac{z}{\sqrt x}\right) - 
2\,\frac{{\cal Q}+\nu_S}{1-\nu_S} \, \frac{\ln\,z}{\ln\, x}\,.
\ee
$G(z)$ has two branch cuts (ignoring for the moment, the logarithmic branch point) and the saddle point equation can be expressed as a gluing condition across these two branch cuts:
\be
G\left(\sqrt x\, z\,\pm\,\epsilon\right)\,=\, G\left(\tfrac{z}{\sqrt x}\,\mp\,\epsilon\right)\,.
\ee
In principle, this condition and the asymptotics of $G(z)$ uniquely determine the resolvent. We will not pursue the analysis further in this paper. Curiously, such a gluing condition defines a Riemann surface of genus one; any significance of this to the physical problem at hand is unclear. Very similar ``gluing problems'' arise as solutions to large-$N$ holomorphic matrix models in very different contexts \cite{kostov}. 

\subsection{Fermion flavour modes}
We now turn to the physics of quarks or fermionic flavour modes on $S^3$. The matrix model incorporating these at finite chemical potential has already been studied in detail in \cite{Hands:2010zp}. 
Technically speaking, the saddle point equations for the fermionic flavours are related to the bosonic ones by the replacement $\zeta \to -\zeta$ and $\nu_S\to -\nu_F$. However, the phase structure is very different, as must be expected for states with different statistics. 

In supersymmetric theories on $S^3$, fermions are heavier than their bosonic counterparts (see Table\eqref{table} in the Appendix). However, if 
the quarks happen to be the lightest modes charged under $U(1)_B$ with energy $\e_0$ and degeneracy $d$, the number of (lightest) baryon states would be
\be
{\mathfrak D}(B)\,=\, \binom{2 N_F\,d + N}{N}\,\approx
\,\exp\left[N\,\ln\left(\frac{(1+d\,\nu_F)^{1+d\,\nu_F}}{(d\,\nu_F)^{d\,\nu_F}}\right)\right].
\ee
Here $N_F$ is the number of Dirac fermion flavours. Now gauge-singlet baryon operators exist for any non-zero value of $N_F$ with degeneracies scaling exponentially with $N$. Consequently, we expect a phase transition when $\mu = \e_0 - \tfrac{T}{N}\ln\,{\mathfrak D}(B)$. Such a phase transition was indeed seen in \cite{Hands:2010zp}. In fact, the fermionic system displays {\em two} such transitions, one on either side of $\mu =\e_0$ at 
\be
\mu^{\mp}_{\rm crit}\,=\,\e_0 \mp\tfrac{T}{N}\,\ln\,{\mathfrak D}(B)\,.
\ee
Both these transitions are third order. The first is a deconfinement transition of the kind we have discussed above, where the baryons ``dissolve'' into quarks and the distribution of eigenvalues of the Polyakov loop matrix acquires a gap. The second transition is less intuitive, but can be understood as a complete occupation of the lowest Fermi level following which the theory re-enters a confined phase. Across this transition the eigenvalues of the Polyakov loop matrix return from the gapped to an ungapped continuum distribution.

Let us briefly  re-derive the above results for fermion flavours, by making use of the replacement $\zeta \to -\zeta$ and $\nu_S \to -\nu_F$ in Eq.\eqref{saddle}. Assuming that the degeneracy of the lightest quark modes $d=1$, we follow the analysis in Section 3.4. The crucial difference with respect to the bosonic case is that the pole (at $Z_\infty =-1/\zeta $) and zero (at $Z_0 = -1/\zeta(1+\nu_F))$ are always on opposite sides of the eigenvalue distribution in the ungapped phase. In addition, the zero always hits the contour ${\cal C}$ with increasing $\zeta$ irrespective of the value of $\nu_F$, thus triggering a GWW transition. This is as expected for fermions, since the corresponding baryon operators $B_F$ exist for any $\nu_F$. Similarly, adapting the formula for baryon number Eq.\eqref{bnumber} to fermions, we find that ${\cal Q}$ increases from ${\cal Q}=0$ to a maximum of ${\cal Q} =\nu_F$ as $\zeta$ is increased from $\zeta_{\rm crit}^- = \nu_F^{\nu_F}/(1+\nu_F)^{1+\nu_F}$ to $\zeta_{\rm crit}^+ =(1+\nu_F)^{1+\nu_F}/\nu_F^{\nu_F}$. Finally we turn to the formula for the end-points of the gapped distribution \eqref{extent} and modify it to suit the fermionic case. We immediately obtain that the the contour is closed, when ${\cal Q}=0$ and also when ${\cal Q}=\nu_F$. These are the two successive GWW transitions.

The main physical effect is that following the two third-order transitions, given a degeneracy $d$ for the lightest fermion harmonic,  baryon number jumps by $d\,\nu_F\,N$. Such jumps (each accompanied by a pair of third order transitions) occur with increasing $\mu$ as  successive fermionic energy levels are encountered and filled. 
We refer the reader to \cite{Hands:2010zp} for further details.

\section{Finite temperature, small $\mu$ regime}

We have explored the low temperature regime of the phase diagram in the $\mu-T$ plane. The remaining question is what happens to the system as the temperature is increased for any fixed $\mu$. We know that the low temperature phase boundary between confined (or mesonic) and deconfined phases is a straight line whose slope is determined by the logarithm of the large-$N$ degeneracy of the (light) baryonic states. For small $\mu$, the finite temperature behaviour can be understood along the lines of  \cite{Aharony:2003sx, Schnitzer:2004qt, Schnitzer:2006xz, Basu:2008uc}. Defining the ``moments'' of the eigenvalue distribution (the normalized Polyakov loops)
\be
\rho_n\, \equiv\, \frac{1}{N}\,\Tr\, U ^n\,, \qquad\rho_{-n}\,\equiv\,\frac{1}{N}\,\Tr\, U^{\dagger n}\,,
\ee
the effective action is
\bea
S_{\rm eff}[\rho]\,=\,&&N^2\,\sum_{n=1}^\infty\frac{1}{n}\,
\left[(1- z_V(x^n)-n_S\, z_S(x^n) + (-1)^n n_F\,z_F(x^n))\,\rho_n\,\rho_{-n}\right.\\\nonumber
&& \left. - \left(\,\nu_S\, Z_S(n\tfrac{\beta}{R},\, m_S R) - \,\nu_F\,(-1)^n\,Z_F(n\tfrac{\beta}{R},\, m_F R)\right)
\left(\rho_n\, e^{n\beta\mu}+e^{-n\beta\mu}\,\rho_{-n}\right) \right]
\eea
The free gauge theory with no flavours ($\nu_S=\nu_F=0$) exhibits a first order deconfinement transition with increasing temperature \cite{Aharony:2003sx}. In pure Yang-Mills theory ($n_S=n_F=0$), this happens when $z_V(x)=1$, and the ``lightest mode'' $\rho_1$ is forced to condense. From this condition, the corresponding Hagedorn/deconfinement temperature is $TR \,=\, 1/\ln(2+\sqrt{3})$. Similarly for ${\cal N}=4$ SYM ($n_S=6\,,n_F=4$), the first moment $\rho_1$ condenses when $z_V(x)+6\,z_S(x)+4\, z_F(x)=1$ or when $TR=1/\ln(7+4\sqrt 3)$. 

In the theory with fundamental flavours the situation is slightly different. For generic non-zero $\mu$, both $\rho_n$ and $\rho_{-n}$ are non-zero and unequal, and their values determined by solving the (linear) equations of motion, 
\be
\rho_n\,=\, e^{-n\beta\mu}\,\frac{\nu_S\, Z_S(n\tfrac{\beta}{R},\, m_S R) - \,\nu_F\,(-1)^n\,Z_F(n\tfrac{\beta}{R},\, m_F R)}{1- z_V(x^n)-n_S\, z_S(x^n) + (-1)^n n_F\,z_F(x^n)}\,
\ee
for $n\neq 0$ and $\rho_0=1$.  The first point to note is that upon plugging the solution back into the action, as long as this saddle point is valid, we expect that the free energy 
is the sum of terms of ${\cal O}(N_S^2)$, ${\cal O}(N_F^2)$ and ${\cal O}(N_F N_S)$. This is indeed the case in the low temperature ungapped phase 
\cite{Schnitzer:2004qt}, leading to the interpretation that the grand potential should be attributed to colour-singlet mesons. This is the confined phase.

 The onset of the deconfinement transition is deduced by locating (for any given $\mu$), the value of the temperature at which a zero appears on the contour ${\cal C}$ of eigenvalues. It is a complicated excercise in general, but somewhat easier for vanishing $\mu$. 
 Setting $\mu=0$ and examining the effective potential for the ``lightest modes'' $\rho_1$ and $\rho_{-1}$ (in an approximation which ignores all higher moments), we obtain,
\be
\rho_{1}\,=\,\rho_{-1}= \frac{\nu_S \,Z_S(\tfrac{\beta}{R}, m_S R) +\nu_F \,Z_F(\tfrac{\beta}{R}, m_F R)}{1-z_V(x)-n_S\, z_S(x) -n_F\,z_F(x)}\,.
\ee
The non-uniform distribution develops a gap when $\rho_1=\rho_{-1}=\frac{1}{2}$, 
so that the spectral density on the circle then takes the form
$\rho(\theta)=\tfrac{1}{2\pi}(1+\cos\theta)$ with a zero at $\theta=\pi$. This is a third order GWW transition \cite{Schnitzer:2004qt}, and is the natural continuation of the low temperature phase boundary to finite temperature and vanishing $\mu$.

\section{The phase diagram: discussion and conclusions}

The general picture is now fairly clear: 
at low enough temperature, with increasing baryon number chemical potential, the theory on $S^3$ is subject to a third order GWW deconfinement transition when the baryon number chemical potential ($N\mu$) approaches the threshold set by the mass of the lightest baryon state and its exponentially large degeneracy in the Veneziano limit. Depending on whether the theory has fundamental fermions or not, there can be more than one such deconfinement transition. When the mass of the lightest scalar baryon (a baryonic operator made purely from fundamental and adjoint scalars) is approached, a third order baryon melting transition ensues followed by the onset of Bose condensation of the squark degrees of freedom. The free gauge theory is not defined beyond this point. 
\begin{figure}[h]
\begin{center}
\epsfig{file=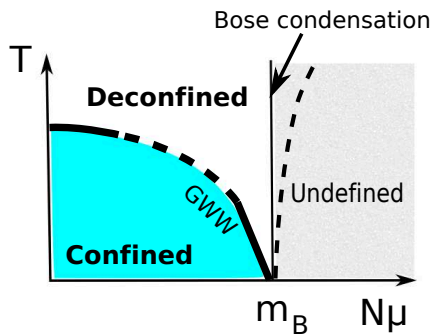, width =2.5in}
\end{center}
\caption{\small{The phase diagram for the free theory, assuming that the lightest baryon states are constituted of squarks. (See Figure(1) for the general situation when quarks are light.)
}} 
 \label{phasediag} 
 \end{figure}
We have also shown that in a different region of the $\mu-T$ plane, namely, low $\mu$ and finite $T$, the theory undergoes a third order transition as a function of temperature. Putting these pieces together we arrive at the general form of the phase diagram depicted in Figures\eqref{phasediag1} and \eqref{phasediag}.  

In both figures, the shaded region with $N\mu > m_B$ is a region of instability for the free theory, where the grand canonical ensemble is ill-defined
(e.g. \cite{yamada, critical}). It is conceivable that upon introducing appropriate self-interaction potentials, the shaded region may have a stable ground state with a VEV for a gauge invariant operator, breaking $U(1)_B$ spontaneously. It can be shown that for SUSY QCD theories or the theory corresponding to the D3-D7 system (${\cal N}=2$ SYM with hypermultiplets), there are flat directions in the scalar potential (Higgs branches and mixed Coulomb-Higgs branches) which are then responsible for runaway unstable directions when $\mu$ is non-zero. For any non-zero temperature and small coupling however, at the origin of the field space, the squarks and adjoint scalars will obtain a temperature dependent thermal mass so that $m_S(T)^2 \sim m_S^2 + \lambda \delta(T)$, where $\lambda$ is the 't Hooft coupling and $\delta(T)$ is exponentially small ($\sim e^{-1/TR}$) at low temperature \cite{critical} and quadratic in $T$ at high temperature. This suggests that, with increasing temperature, the chemical potential will need to be increased to a higher threshold value before a runaway instability kicks in. Therefore we may expect a 
small region in the $\mu-T$ plane, bounded by $N\mu=m_B$ and the dashed line in Figures\eqref{phasediag1} and \eqref{phasediag}, where a metastable, deconfined phase exists and where all scalars have vanishing expectation values (see \cite{yamada, critical} for analogous discussions in the case of ${\cal N}=4$ SYM with R-symmetry chemical potentials). 

At strong coupling (and in flat space), the unstable region has been studied in the context of the D3-D7 system for a small number of flavours 
\cite{Chen:2009kx} and it should be possible to investigate this within the framework of \cite{Bigazzi:2011it} which incorporates the back-reaction from flavour branes. In the `probe' approximation (fixed number of flavours in the large-$N$ limit), the phase diagram in the $\mu-T$ plane of the D3-D7 system was extensively explored in \cite{mateos, mateos1,obannon}. Despite being in the probe limit, we would expect a Bose condensed phase (in addition to a Fermi surface from the quarks) due to squark modes charged under $U(1)_B$ in the setup (see \cite{mateos, obannon, Ammon:2011hz} for related discussions). An extensive search for instabilities and $U(1)_B$ breaking in the world-volume fluctuations of a probe D7-brane in $AdS_5\times S^5$ has recently been carried out in \cite{Ammon:2011hz}, with a negative result. 
It would be interesting to understand this better within a general setup of the kind we have studied but with (small) interactions switched on. It would also be of interest to find what role, if any, is played by states such as adjoint-baryons in this picture.

The low temperature, finite density transitions we have found in the Veneziano large-$N$ limit, are due to the exponentially large degeneracy of baryon states. It is reasonable to expect these to persist for any coupling, at large-$N$ (e.g \cite{Hidaka:2008yy}). If so, within an appropriate dual gravity set-up, this would correspond to a black hole solution dual to the phase of cold, dense, deconfined (s)quarks. 

A final comment should be made about the phases that we have found (Section 4.2) with order parameters that involve modes of the fields that carry non-trivial angular momentum . These inhomogeneous phases should involve the spontaneous breaking of rotational symmetry on the $S^3$. This feature would be worthy of future study.
\\
\\
{\bf Acknowledgements:} We would like to thank Paolo Benincasa, Aldo Cotrone, Tom DeGrand, Maciej Nowak and Andy O' Bannon for useful discussions. We especially thank Shiraz Minwalla and Sameer Murthy for an intense, stimulating discussion which led to new insights.
SPK and JCM would like to thank the
Galileo Galilei Institute for Theoretical Physics for the hospitality and the opportunity to
present and discuss parts of this research, and the INFN for partial support during the ``Large-$N$'' workshop, 2011. SPK also thanks
the Centro de Ciencias de Benasque Pedro Pascual and the organizers of the Gravity workshop, July 2011 during which parts of this work were completed and presented.
The work of JCM is supported by the Stichting voor Fundamenteel Onderzoek der Materie (FOM). TJH and SPK are supported by STFC Rolling Grant ST/G0005006/1.

\newpage
\startappendix
\Appendix{Harmonics on $S^3$}
\label{sec:appendix}
\begin{table}[ht]
\caption{Harmonics on  the three-sphere}
\center
\begin{tabular}{ c c c c} 
\hline
\hline
Field & Angular momentum ($\ell$) & Energy ($\e_\ell$) & Degeneracy ($d_\ell$) \\ [0.5ex]
\hline
$B_i$ & $\ell >0$ & $(\ell+1) R^{-1}$ & $2\ell(\ell+2)$ \\[0.5ex] 
$C_i$ & $\ell >0$ & $\sqrt{\ell(\ell+2)}R^{-1}$ & $(\ell+1)^2$ \\[0.5ex] 
$A_0$ & $\ell \geq 0$ & $\sqrt{\ell(\ell+2)}R^{-1}$ & $(\ell+1)^2$ \\[0.5ex] 
Fermions & $\ell>0$ & $(\ell+\tfrac{1}{2}) R^{-1}$ & $2\ell(\ell+1)$ \\[0.5ex] 
Massless scalars & $\ell \geq0$ & $(\ell+1)R^{-1}$ &  $(\ell+1)^2$ \\
(conformal coupling) & & & \\
\hline
\end{tabular}
\label{table}
\end{table}

We list the energies of Kaluza-Klein harmonics of scalar, vector and
fermion fields on a spatial three-sphere. For details of the spherical harmonic decomposition we refer the reader to the discussion in \cite{sbh}. The main point to note is that fermions $\psi$, and the spatial modes of the vector field denoted by $A_i$ ($i=1,2,3$), do not have zero modes on $S^3$. In general, the spatial components of a gauge field $A_i$ can be split  into the image and kernel of $\nabla_i$ (the former are the transverse, physical degrees of freedom). In particular $A^i=B^i+C^i$ with $C^i=\nabla^i f$ and $\nabla_i B^i=0$. The time component of the gauge field $A^0$ and scalar fields are allowed to have zero modes on $S^3$.

\end{document}